\documentclass[a4paper,fleqn,usenatbib]{mnras}

\usepackage{newtxtext,newtxmath}

\usepackage[T1]{fontenc}
\usepackage{ae,aecompl}
\usepackage{subfigure}
\usepackage{multirow}
\usepackage{placeins}

\usepackage{graphicx}	
\usepackage{amsmath}	
\usepackage{amssymb}	
\usepackage[color=yellow]{todonotes}
\usepackage{times}

\newcommand{\xray}{V410~X-ray~1}	

\title[\xray]{The extremely truncated circumstellar disc of \xray: \newline a precursor to TRAPPIST-1?}

\author[D.~M.~Boneberg et al.]
{\parbox{\textwidth}{D.~M.~Boneberg$^{1}$\thanks{E-mail: boneberg@ast.cam.ac.uk},
		S.~Facchini$^{2}$,
		C.~J.~Clarke$^{1}$,
		J.~D.~Ilee$^{1}$,
		R.~A.~Booth$^{1}$
		and S.~Bruderer$^{2}$
	}\vspace{0.4cm}\\
	\parbox{\textwidth}{
		$^{1}$Institute of Astronomy, Madingley Road, Cambridge CB3 0HA, UK\\
		$^{2}$Max-Planck-Institut f\"{u}r Extraterrestrische Physik, Giessenbachstrasse 1, D-85748 Garching, Germany
	}}

	\date{Accepted XXX. Received YYY; in original form ZZZ}
	
	\pubyear{2018}
	
\begin{document}
		\label{firstpage}
		\pagerange{\pageref{firstpage}--\pageref{lastpage}}
		\maketitle
		
		\begin{abstract}
			Protoplanetary discs around brown dwarfs and very low mass stars offer some of the best prospects for forming Earth-sized planets in their habitable zones.  To this end, we study the nature of the disc around the very low mass star \xray, whose SED is indicative of an optically thick and very truncated dust disc, with our modelling suggesting an outer radius of only 0.6\,au.  We investigate two scenarios that could lead to such a truncation, and find that the observed SED is compatible with both. The first scenario involves the truncation of both the dust and gas in the disc, perhaps due to a previous dynamical interaction or the presence of an undetected companion.  The second scenario involves the fact that a radial location of 0.6\,au is close to the expected location of the H$_2$O snowline in the disc.  As such, a combination of efficient dust growth, radial migration, and subsequent fragmentation within the snowline leads to an optically thick inner dust disc and larger, optically thin outer dust disc.  We find that a firm measurement of the CO $J=2$--1 line flux would enable us to distinguish between these two scenarios, by enabling a measurement of the radial extent of gas in the disc.  Many models we consider contain at least several Earth-masses of dust interior to 0.6\,au, suggesting that \xray\ could be a precursor to a system with tightly-packed inner planets, such as TRAPPIST-1.
		\end{abstract}
		
		\begin{keywords}
			stars: pre-main sequence -- planetary systems: protoplanetary discs -- circumstellar matter -- 
			stars: brown dwarfs  -- techniques: interferometric 
		\end{keywords}
		
		
		
		\section{Introduction}
		\label{sec:introduction}
		
		A number of theories have been proposed for the formation channel of brown dwarfs (BDs) and very low mass (VLM) stars \protect\citep[see e.g.][]{WhitworthEtAl2007}. These objects are distinguished according to their mass, where a BD falls below the hydrogen-burning mass limit of $\sim 0.08\,$M$_\odot$ \protect\citep{OppenheimerEtAl2000} and a VLM object would lie above this mass boundary.  The formation scenarios of these objects can be broadly divided into two categories.  Firstly, BD formation may be a scaled-down version of ordinary (solar mass range) star formation, involving the collapse of an essentially isolated low mass gas core.  Secondly, a range of other scenarios exist in which the BD properties are impacted by their formation environment.  The latter includes dynamical influences, such as the ejection of stellar embryos from systems with a small number of stars \protect\citep{ReipurthClarke2001} and the possibility of BD formation and subsequent ejection from the outer regions of massive circumstellar discs \protect\citep{StamatellosWhitworth2009}.
		
		\smallskip
		
		Perhaps the most important observational discriminant is the radii of discs around BD and VLM stellar objects. Ejections resulting from dynamical encounters are expected to truncate circumstellar discs at a radius of around a third of the distance of closest approach \protect\citep[e.g.][]{HallEtAl1996, BreslauEtAl2014,WinterEtAl2018}.  Massive objects tend to remain within their natal gas reservoirs after encounters and can thus re-accrete disc material. Ejected objects (which are preferentially low mass) however bear the imprint of their last dynamical encounter in the form of a truncated disc. Given the stochastic nature of dynamical interactions, the range of closest approaches for BDs in dynamical simulations is very large (around three orders of magnitude); moreover, in the simulations of \protect\citet{Bate2009} around a half of all BDs have suffered encounters on a scale of a few au.   It is thus a prediction of dynamical star formation scenarios that BD or VLM stars should have a much larger range of disc radii than higher mass stars, and that some population of extremely compact discs is to be expected.  Companions can also leave an imprint on the disc size. It is expected that for a binary system the disc radius will be truncated at about 20-50 per cent of the semi-major axis of the binary \protect\citep{ArtymowiczLubow1994}.
		
		\smallskip
		
		There is increasing interest in the planet formation potential of discs in the BD and VLM star regime.  This is partly because of the growing realisation that lower mass host stars provide the best prospects for the detection of Earth-like planets in the habitable zone.  Recent results have borne out this expectation, in particular the discovery of 7 terrestrial planets within 0.06\,au around the 0.08\,M$_\odot$ star TRAPPIST-1 \protect\citep{GillonEtAL2017}. Such systems imply an extraordinary concentration of solid material at small orbital radii and raise a number of unanswered questions about the evolutionary scenario that produced them.  In particular, it is of obvious interest to ask what would be the distribution of dust and gas in such a system at an age of a few million years?
		
		\smallskip
		
		To date, a number of studies have identified and characterised the population of young BDs and VLM objects which have large discs ($>70$\,au; \protect\citealt{RicciEtAl2013,RicciEtAl2014}). The large discs have been imaged with the Atacama Large Millimeter/submillimeter Array (ALMA) and the Combined Array for Research in Millimeter-wave Astronomy (CARMA), and have broadband SEDs which are consistent with extended circumstellar discs \protect\citep{AlvesdeOliveiraEtAl2013}.  Additionally, their infrared colours in the 2--12\,$\mu$m and 12--70\,$\mu$m range lie in the domain of Classical T Tauri stars \protect\citep{RodgersLeeEtAl2014}.  \protect\cite{VanDerPlasEtAl2016} surveyed 8 brown dwarf discs in Upper Scorpius and Ophiuchus with ALMA, which, being unresolved in these observations, limits the discs to being $R\lesssim40\,$au. 
		Recently, \protect\cite{TestiEtAl2016} discovered discs around BDs in the $\rho$ Oph star-forming region with ALMA that have sharp outer edges at radii of 25 au. 
		
		\smallskip
		
		However, there seems to be a population of even smaller discs: \protect\cite{BulgerEtAl2014} present spectral energy distributions (SEDs) of a large sample of low mass members of Taurus, and mention the possibility of a `truncated disc' population, based on a steeply declining SED from $\sim$20$\micron$ appearing to follow the Rayleigh-Jeans limit. Hints for small discs around VLM Objects and BDs have also been found by \protect\cite{HendlerEtAl2017} using  $63\,\micron$ continuum and [O I] observations from the PACS spectrometer in the Taurus and Chamaeleon I regions. Assuming disc geometry and dust properties based on T Tauri stars, they obtain from their SED modelling discs with radii between $\sim1-80\,$au. However they stress that further (spatially resolved) ALMA observations are needed to confirm this.
		
		\smallskip
		
		Recent studies have attempted to determine the amount and properties of dust in discs around BDs or VLM stars. These can give important hints as to the evolution of the disc, but also the propensity for the disc to form planets. \citep{Ward-Duong_2018} modelled 24 BD/VLM objects in Taurus, finding a range of dust masses in the range 0.3--20\,M$_{\oplus}$ and an approximately linear relationship between the stellar mass and dust mass in these discs.  BD discs are expected to have a smaller population of mm-size grains in comparison to discs around higher mass T Tauri stars because the radial drift velocities of these grains are higher. However, \protect\cite{PinillaEtAl2017} have examined the 3\,mm continuum fluxes of three BD discs in the Taurus star forming region with the IRAM/Plateau de Bure Interferometer combined with previous studies of the 0.89\,mm fluxes obtained with ALMA. They find that from millimetre spectral indices, large grains actually seem to be present in these discs.
		
		\smallskip
		
		\protect\cite{GreenwoodEtAl2017} have recently performed thermochemical modelling of BD discs including predictions for future ALMA observations of molecular tracers such as CO, HCN and HCO$^+$. Their models suggest that BD discs are similar to T Tauri discs and that similar (thermochemical) diagnostics can be used.  Interestingly, \protect\cite{BayoEtAl2017} find from recent ALMA Band 6 continuum observations of the disc around the even lower-mass object OTS44 ($M\sim6-17M_\text{jup}$) that it also possesses a disc and even falls onto a power-law relation between stellar mass and dust disc mass.  
		
		\smallskip
		
		Two broad explanations for truncated discs exist.  Firstly, a dramatic segregation of dust and gas may occur due to the radial migration of millimetre-sized dust within what could be termed a `normal' sized gas disc.  Indeed, simulations of dust particles in BD discs have shown that migration can be more significant than in T Tauri discs \citep{PinillaEtAl2013}.  Secondly, the small radial extent may be shared by both the dust and the gas in the disc.  Such a scenario would imply a dynamical origin for the small radial extent.  In order to determine which of these scenarios may be causing the observed population of truncated BD discs, it is crucial to examine both the dust and gas components of the discs. 
		
		\smallskip
		
		In this paper, we present observations and detailed modelling of a candidate VLM star possessing a very truncated disc -- \xray.  We describe the observational data that goes into the modelling in Section~\ref{sec:observations} and present the thermochemical modelling with DALI in Section~\ref{sec:methods}. The results of the modelling are then given in Section~\ref{sec:results}, where we also discuss two possible explanations for the truncation of the dust disc of \xray. In Section~\ref{sec:conclusion}, we summarise the findings and present the main conclusions of this work.

		\begin{figure}
			\centering
			\includegraphics[width=\columnwidth]{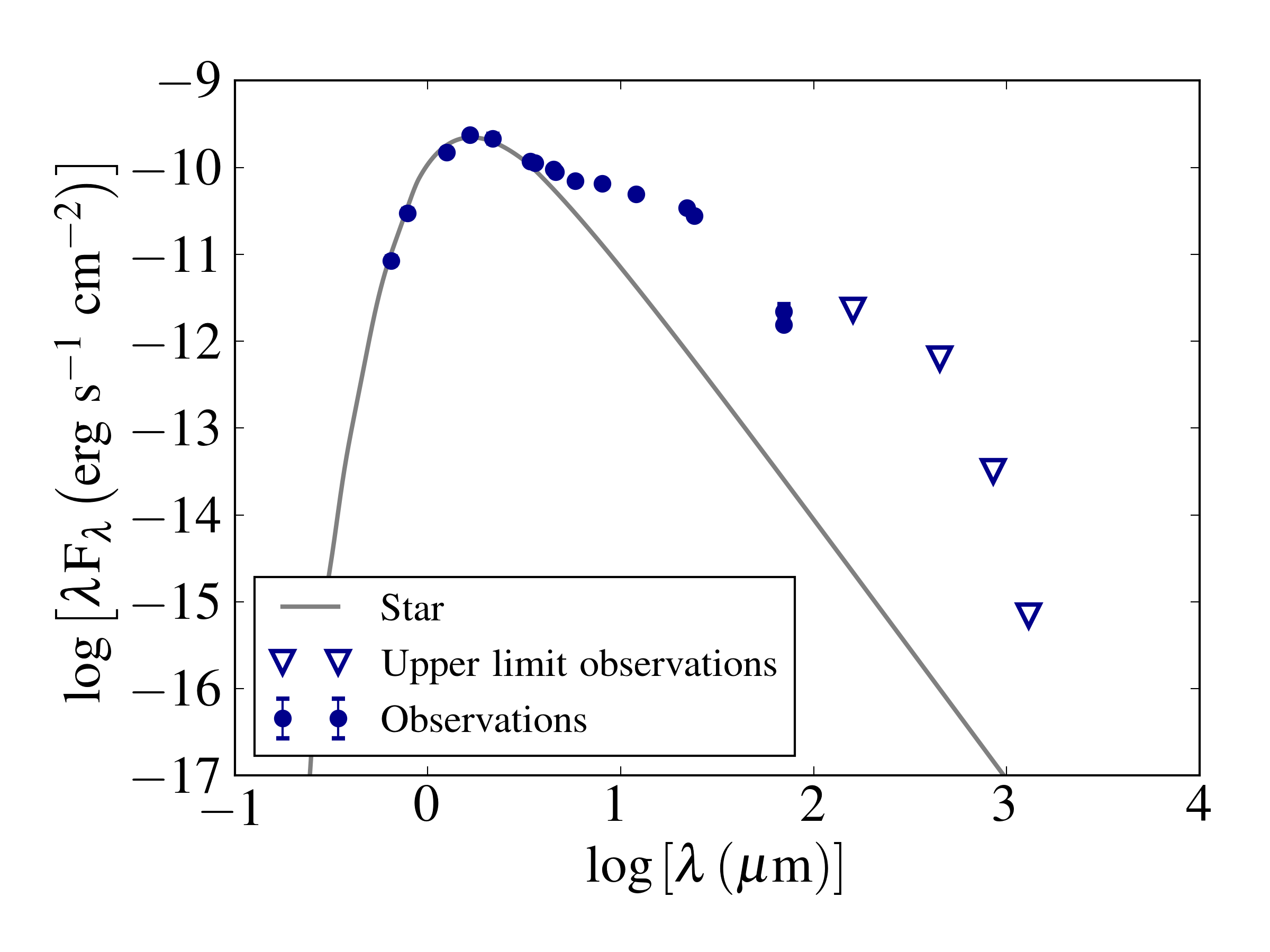}
			\caption{SED of \xray. The fluxes and corresponding references are listed in Table~\ref{tab:sed}. Upper limits are given by triangles. The limit we derive on the continuum flux from ALMA observations is given by the blue triangle at $\lambda=1.3\,$mm. The error bars are shown, but are of the order of the size of the data
				points.}
			\label{fig:SED}
		\end{figure}

		\section{Observations}
		\label{sec:observations}
		
		\xray\ is a very low mass star located within the Taurus star forming region L1495 at a distance of $d\sim140$\,pc, lying at co-ordinates RA$=04^h17^m49\mbox{\ensuremath{.\!\!^{s}}}655$, Dec.$=+28^\circ 29'36\mbox{\ensuremath{.\!\!^{\prime\prime}}}27$ (J2000).  It has a spectral type determined to be M4 \citep{AndrewsEtAl2013} or M2.6 \citep{herczeg2014}.   In order to compile a spectral energy distribution (SED) of \xray, we draw on photometric measurements from the literature from multiple instruments spanning a range of wavelengths from the optical to the sub-millimetre.  The wavelengths, fluxes and appropriate references for these data are given in Table \ref{tab:sed}. 
		
		\smallskip
		
		In addition to the published fluxes, we also obtained archival ALMA observations of \xray\ \protect\citep{SimonEtAl2017}.  The observations were taken on 2015-09-19 in Band 6 (230\,GHz, 1.3mm) and the on-time source was 6 minutes.  We pipeline reduce the data, giving a synthesised beam of $0.26 \times 0.21\arcsec$ (or $36 \times 29$\,au at a distance of 140\,pc).  In addition to the continuum, the observations also covered the $^{12}$CO $J=2$--1 transition with a velocity resolution of 0.2\,km~s$^{-1}$.  Both the continuum and line observations with ALMA resulted in a non-detection.  Based on these non-detections, we calculate a 3$\sigma$ upper limit for the continuum flux at 1.3\,mm to be 0.3\,mJy, and a 3$\sigma$ upper limit for the $^{12}$CO $J=2$--1 line flux to be 0.4\,Jy\,km\,s$^{-1}$.  The latter is based on the assumption of a line width of 10\,km\,s$^{-1}$, appropriate for the conditions expected in this disc\footnote{This velocity corresponds roughly to few au radius around a 0.1$M_\odot$ star. The line width is set by the inner radius of the CO emission, which is usually much further out than the inner disc radius (0.04\,au in our case). The modelling of the CO emission as presented later in the paper does indeed confirm that the inner radius of the CO emission is further out, such that a line width of 10\,km\,s$^{-1}$ should be wide enough to comprise the entire emission. We note that even if this estimate is slightly wrong, it only enters the calculation of the flux with a square root dependence.}.  We include the 1.3\,mm upper limit in Table \ref{tab:sed} and Figure \ref{fig:SED}, and use the $^{12}$CO $J=2$--1 upper limit in our analysis in Section \ref{sec:results}.

		\begin{table}
			\centering
			\begin{minipage}{75mm}
				\begin{center}
					\bgroup
					\caption{Wavelengths ($\lambda$) and associated flux densities (F) used for compilation of the SED of \xray.}   
					\label{tab:sed} 
					\begin{tabular}{llll}
						\hline
						$\lambda$ ($\mu m$) & F (mJy) & Band & Reference \\
						\hline
						0.65    & 1.8     & Rc            &\citet{BulgerEtAl2014} \\ 
						0.79    & 7.8     & Ic            &\citet{BulgerEtAl2014} \\
						1.25    & 62      & 2MASS J             &\citet{CutriEtAl2003}  \\
						1.65    & 131       & 2MASS H             &\citet{CutriEtAl2003}  \\
						2.17    & 155       & 2MASS Ks            &\citet{CutriEtAl2003}  \\
						3.4    & 133       & WISE W1       &\citet{WrightEtAL2010} \\
						3.6     & 136       & IRAC 3.6      &\cite{LuhmanEtAl2010}  \\
						4.5     & 143       & IRAC 4.5      &\cite{LuhmanEtAl2010}  \\
						4.6     & 136        & WISE W2       &\citet{WrightEtAL2010} \\
						5.8     & 136        & IRAC 5.8      &\citet{LuhmanEtAl2010} \\
						8.0     & 175       & IRAC 8.0      &\citet{LuhmanEtAl2010} \\
						12    & 196       & WISE W3       &\citet{WrightEtAL2010} \\
						22    & 253       & WISE W4       &\citet{WrightEtAL2010} \\
						24    & 221       & MIPS 24       &\citet{RebullEtAL2010} \\
						70    & 36      & PACS 70       &\citet{RebullEtAL2010} \\
						70    & 51      & MIPS 70       &\citet{RebullEtAL2010} \\
						160   & $<$122    & PACS 160      &\citet{RebullEtAL2010} \\
						450   & $<$94     & SCUBA-2       &\citet{MohantyEtAl2013} \\
						850   & $<$9      & SCUBA-2       &\citet{MohantyEtAl2013} \\
						1300  & $<$0.3    & ALMA B6 & This work \\
						\hline
					\end{tabular}
					\egroup
				\end{center}
			\end{minipage}
		\end{table}

		\section{Methods}
		\label{sec:methods}
		In this paper  we explore  various scenarios for the circumstellar environment of \xray\ that are consistent with
		the  SED shown in Figure~\ref{fig:SED} and with the upper limits on the
		$^{12}$CO $J = 2$--1 flux.

		To this end we use the radiation thermo-chemical disc code DALI (Dust And LInes, \citealt{BrudererEtAl2012,Bruderer2013}) with adaptations as detailed below to model the gas and dust emission in \xray. DALI solves the continuum radiative transfer equations to obtain the dust temperature, and the thermal balance and chemical abundances to compute the gas temperature structure. The ray tracing module is then used to obtain both an SED and CO fluxes that can be compared against the observations. 
		
		\smallskip
		
		In DALI, gas and dust distributions that are not necessarily co-spatial can be taken into account. We use two dust grain populations with small grains being coupled to the gas distribution and large grains that are settled with respect to the gas. For the study described in Section \ref{subsec:extent}, we use size ranges for small and large grains of 5\,nm-10\,$\mu$m and 10\,$\mu$m--0.3\,mm respectively. We use a size distribution index $q=3$ for the distribution of grain sizes and employ a mass ratio of large to small grains of 30. We discuss our parameter choices for a model with spatially variable maximum grain size in Section \ref{subsec:origin_migration}. 
		The large grains are settled with respect to the gas such that their scaleheight is reduced to 0.2$h_\text{gas}$. The dust opacities are taken from the opacity library used by \citet{2017A&A...605A..16F,2017arXiv171004418F}. More specifically, opacities are computed from Mie theory using the \verb!miex! code \citep{2004CoPhC.162..113W}. Optical constants are taken from \citet{2003ApJ...598.1017D} for graphite and \citet{2001ApJ...548..296W} for silicates.

		We have adapted the code such that the gas is in vertical hydrostatic equilibrium, which in turns therefore also influences the vertical distribution of the dust.  In the  radial direction, the gas surface density profile is set up as follows:
		\begin{equation}
			\Sigma_\text{gas}= \Sigma_\text{c} \left(\frac{r}{R_\text{c}}\right)^{-\gamma} \exp\left[-\left(\frac{r}{R_\text{c}} \right)^{2-\gamma} \right]\hspace{2pt},
			\label{eq:Sigmagas}
		\end{equation}
		where we use $\gamma=0.8$.  When we refer to the gas radius of the models we adopt  $R_\text{gas}\approx 3R_\text{c}$, since this bounds   the region from which typically  more than 90 per cent of the emission arises. 
		
		\smallskip
		
		In order to obtain vertical hydrostatic equilibrium in the gas, we perform the following steps:
		
		\begin{enumerate}
			
			\item For the model in step 1, we guess the scale height as a function of R and prescribe it with the DALI parameters for the flaring angle $h=h_\text{c}(R/R_\text{c})^\psi$, where $\psi$ and $h_\text{c}$ are chosen such that the model SED roughly matches the observed SED. We then calculate the corresponding density distribution $\rho_1(r,z)$ assuming the surface density distribution given in Equation~\ref{eq:Sigmagas} and a Gaussian density dependence on z with the prescribed scale height $h(r)$. DALI is then run to obtain the initial thermal equilibrium temperature distribution $T_1(r,z)$.
			
			\item The density distribution is then recalculated ($\rho_2(r,z)$) as a local Gaussian with its scale height now given by that predicted in hydrostatic equilibrium if the disc was vertically isothermal with $T(r)=T_1(r,z=0)$. DALI is then run to obtain the thermal equilibrium temperature distribution, $T_2(r,z)$, with the new density profile $\rho_2(r,z)$.
			
			\item In the final step, the density distribution is recalculated ($\rho_3(r,z)$) as the density profile that is in vertical hydrostatic equilibrium given $T_2(r,z)$. Note that this density distribution is no longer necessarily Gaussian because the disc is no longer vertically isothermal. DALI is then run a third time to obtain the thermal equilibrium temperature distribution $T_3(r,z)$ corresponding to $\rho_3(r,z)$.
		\end{enumerate}
		
		As a consistency check, $\rho_3(r,z)$ is compared with $\rho_4(r,z)$, which is the density profile that is in hydrostatic equilibrium given $T_3(r,z)$. It is found that the differences between $\rho_3(r,z)$ and $\rho_4(r,z)$ are negligible (except at extremely low densities), and thus the final density and temperature profiles are in both thermal and hydrostatic equilibrium.  Observational diagnostics are derived from the models with $T_3(r,z)$ and $\rho_3(r,z)$, which are consistently reddened for comparison with the real observations assuming A$_{v} = 3.8$ and R$_{v} = 3.1$. 
		
		\smallskip
		
		The stellar parameters are kept fixed and are given in Table~\ref{tab:prop}. We have slightly adapted the stellar parameters given by \protect\cite{BulgerEtAl2014} in order to provide a better match to the stellar part of the SED. They are compatible with the values quoted in \protect\cite{AndrewsEtAl2013}.  As the inclination of the disc is unknown, we use an inclination of $i=45^{\circ}$ for our fiducial model.However, we note that our choice of inclination does not affect the results except for highly inclined discs. The accretion luminosity of \xray\ is unknown, and as such we are unable to assign a mass accretion rate in our models.  However, we have verified that accretion rates as high as $\dot{M} \approx 10^{-8}\,$M$_\odot$~yr$^{-1}$ have no discernible effect on the appearance of the SED\footnote{In this extreme case the CO fluxes can become slightly higher (by a factor of $\sim2$) for large discs. However for more realistic lower accretion rates, the difference will then be much lower.}, and values above this would lie far outside the measured accretion rates for such low mass objects \citep[see, e.g.,][]{Herczeg2009}. \protect\cite{ManaraEtAl2015} even find that typical accretion rates of BDs in young star forming regions are of the order $<10^{-9.5}$M$_\odot$~yr$^{-1}$.
		We use an X-ray luminosity of $L_\text{x}=10^{28}\,$erg~s$^{-1}$ based on \protect\cite{StromStrom1994}. The inner radius of the disc (in gas and dust) is set to $R_\text{in}=0.04$\,au,  based on a calculation of the dust sublimation radius around a star with the parameters given in Table~\ref{tab:prop} and assuming a dust sublimation temperature $T_\text{subl}\sim1500\,$K \citep{WoodEtAl2002,MonnierMillanGabet2002}.
		
		\begin{table}
			\centering
			\begin{minipage}{50mm}
				\begin{center}
					\bgroup
					\caption {Parameters adopted for both the central star and those that are kept constant during the disc modelling procedure.}
					\label{tab:prop} 
					\begin{tabular}{|l|c|}
						\hline
						\bf{Stellar properties}  &  \\
						\hline
						$M_\text{star}$ (M$_\odot$)& 0.1$^{\dag}$ \\ 
						$L_\text{star}$ (L$_\odot$)& 0.4$^{\phantom{x}}$  \\ 
						$T_\text{eff}$ (K)& 3000  \\
						$R_\text{star}$ (R$_\odot$)& 2.3$^{\phantom{x}}$ \\ 
						A$_{\rm V}$ (mag) & 3.8$^{\phantom{x}}$ \\ 
						\hline
						\bf{Disc parameters}  & \\
						\hline
						$R_\text{dust,1}$ (au) & 0.6  \\
						Inclination $i$ ($^{\circ}$) & 45\\
						Inner disc radius $R_\text{in}$ (au) &  0.04\\
						\hline
					\end{tabular}
					\\
					\small{$\dag$: based on \protect{\cite{DantonaMazzitelli1997,AndrewsEtAl2013}}}
					\egroup
				\end{center}
			\end{minipage}
		\end{table}
		
		\section{Results \& Discussion}
		\label{sec:results}
		
		\subsection{Constraining the radial extent of the dust via  SED fitting}
		\label{subsec:extent}
		
		The dust disc radius is   well constrained by the  SED morphology in the Herschel bands.   The slope of $-3$ in the $\lambda F_\lambda, \lambda$ plane between $24$ and $70\,\mu$m suggests that the SED is here dominated by the Rayleigh Jeans tail of the coolest disc material, while the location of the spectral steepening (between $10$ and $24 \mu$m) implies that this corresponds to dust at around $150$\,K.  This is suggestive of the spectrum expected from a truncated optically thick disc, consistent with the designation of \xray\ as a member of the `truncated disc' class identified by \protect\cite{BulgerEtAl2014}.
		Further constraints on the properties of the dust and gas require dedicated modelling, which we perform here.
		
		\smallskip
		
		We have explored the range of dust properties that are compatible with Figure~\ref{fig:SED} and find that, in line with the simple argument above, the SED is best reproduced by models that are optically thick out to a radius where the effective temperature is  $\sim140\,$K (consistent with the onset of the Rayleigh Jeans tail at $24\,\mu$m) and devoid of dust at larger radii. Naturally, we cannot rule out small quantities of optically thin dust beyond this truncation radius (and quantify, in Section~\ref{subsec:origin_migration} just how much dust can be accommodated at larger radius in the context of a specific physical model). Nevertheless, trace dust at large radius is not {\it required} to reproduce the SED.
		
		\smallskip
		
\begin{figure}
			\centering
			\includegraphics[width=\columnwidth]{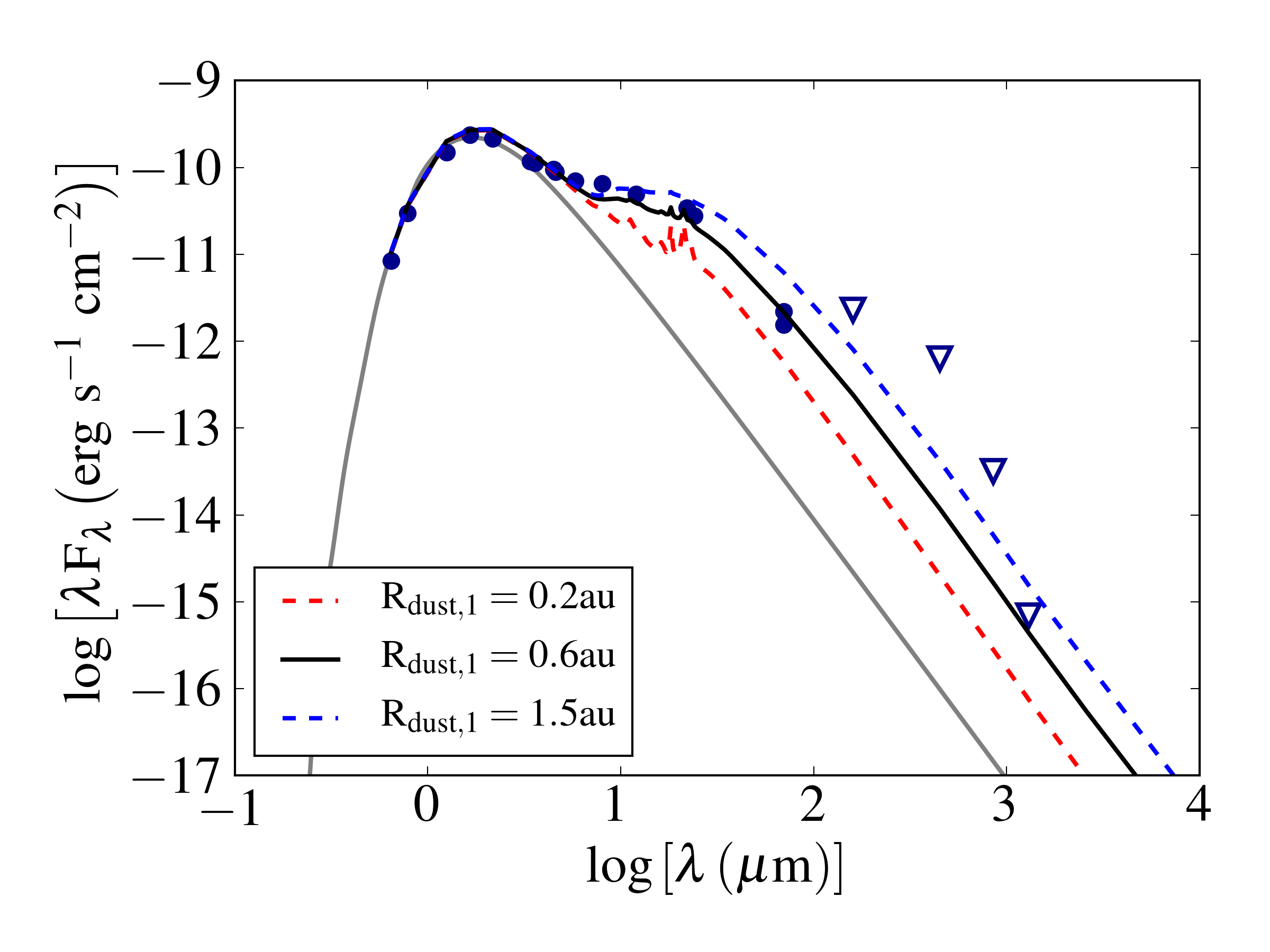}
			\caption{SED of models with various R$_\text{dust,1}$. The model with R$_\text{dust,1}=0.6\,$au corresponds to our `truncation' scenario.  The observational data are identical to those presented in Figure~\ref{fig:SED}.  }
			\label{fig:SED_Rdust}
		\end{figure}

		The  stringent requirement is instead  that the spectrum follows the Rayleigh-Jeans slope at wavelengths beyond $24 \mu$m.  In practice, given the luminosity of the star, our radiative transfer modelling implies that the outer radius is at $R_\text{dust, 1}\sim 0.6\,$au -- Figure~\ref{fig:SED_Rdust} demonstrates the effect of three different truncation radii for discs that are optically thick everywhere at 70\,$\mu$m.
		Despite the non-detection of the flux at 1.3\,mm the radius determined from the SED is remarkably well constrained due to the relatively high sensitivity of the ALMA observations. The two measurements of the 70\,$\mu$m flux from PACS and MIPS along with the uncertainty on the disc inclination lead to a relatively modest uncertainty in the radius of approximately 10 per cent.
		Clearly, larger (or smaller) truncation radii cause the $70\,\micron$ flux to be over (or under-) predicted.  
		We will henceforth describe this model as the `truncation' scenario.
		
		\smallskip
		
		Constraints on the required dust {\it mass} are instead weak and imply only a total dust mass of $M_\text{dust}\gtrsim0.01 M_\oplus$ in order for the disc to be optically thick at $70$ microns.  However, for models with dust masses significantly below $\sim0.1 M_\oplus$, the fluxes in the NIR become too low in comparison with the observations.  For a disc truncated at 0.6\,au the disc is optically thick out to 1.3\,mm for disc dust mass in excess of $\sim1 M_\oplus$ and thus models with dust mass higher than this limit will share the SED shown in black in Figure~\ref{fig:SED_Rdust} (whose flux at 1.3\,mm is slightly below the $3 \sigma$ upper limit provided by ALMA). This implies that dust masses higher than  $\sim1 M_\oplus$ can be present within the innermost 0.6\,au. 
		
		\smallskip
		
		Such a large budget of raw planet-forming material contained within such a relatively small radius suggests that \xray\ may be a precursor of one of the many systems of tightly packed inner planets (STIPs) that have recently been discovered \citep[see, e.g.,][]{Lissauer2011, Fabrycky2014}.  Perhaps the most well known example of a compact system of planets is TRAPPIST-1, which contains at least 7 Earth-sized planets within 0.06\,au of the central star \citep{GillonEtAL2017}.  The combined mass of the planets discovered so far in TRAPPIST-1 (what might be termed a `minimum mass TRAPPIST-1 nebula') is 5.28$M_\oplus$.  Our analysis of the SED cannot rule out the presence of a comparable amount of dust within the inner disc of \xray, suggesting that it may in fact be a precursor to a similarly compact system of terrestrial planets.

		\begin{figure*}
			\centering
			\includegraphics[width=0.4\textwidth]{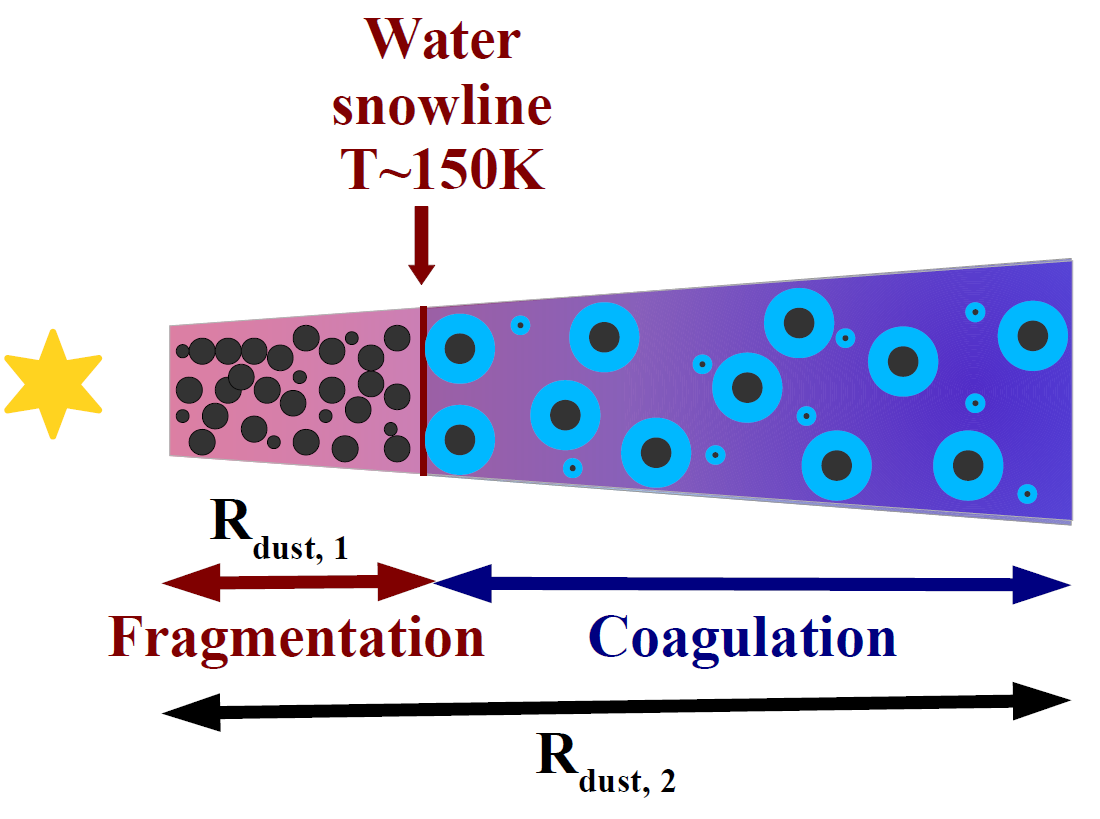} \\
			\includegraphics[width=0.49\textwidth]{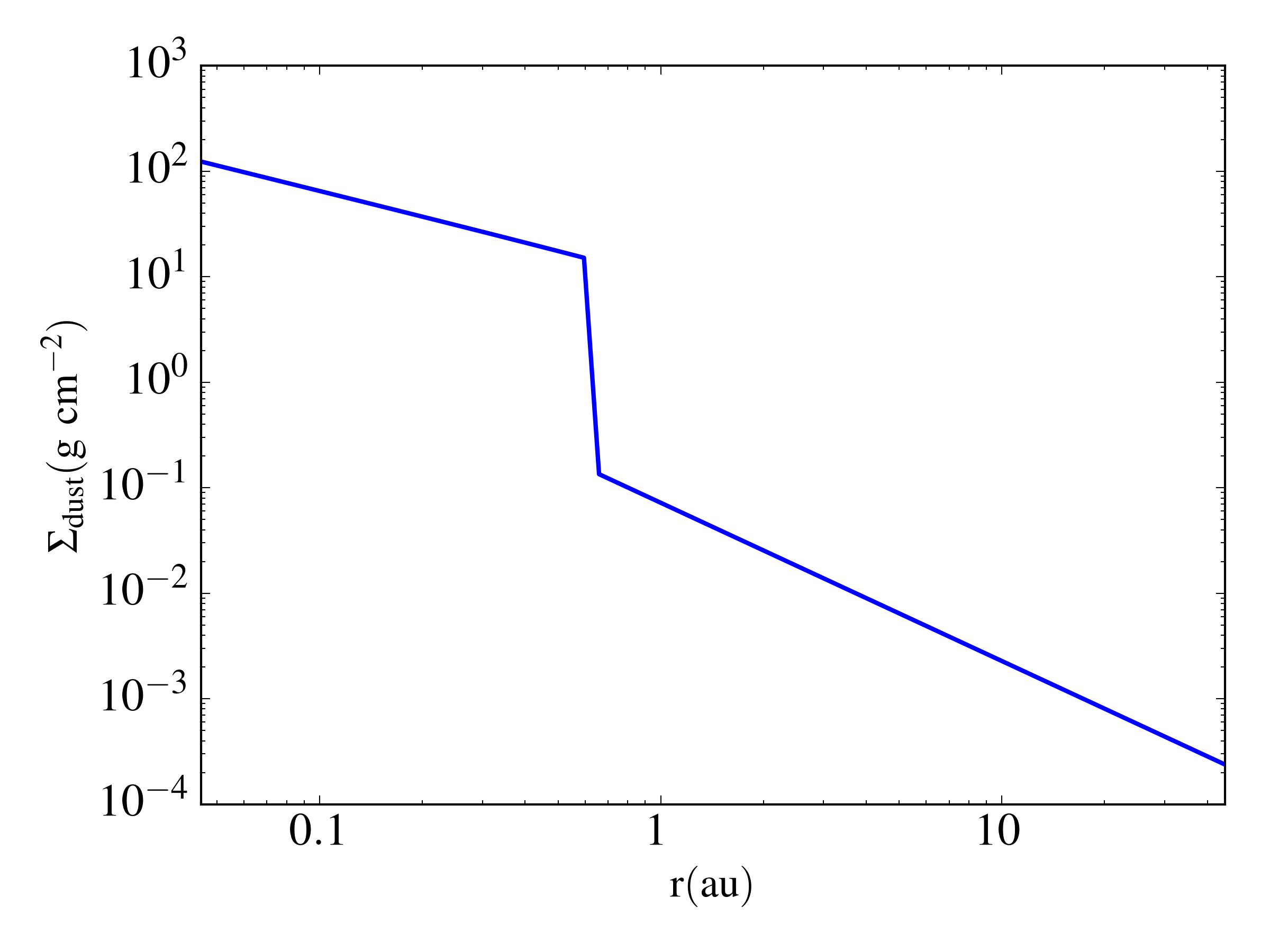}
			\includegraphics[width=0.49\textwidth]{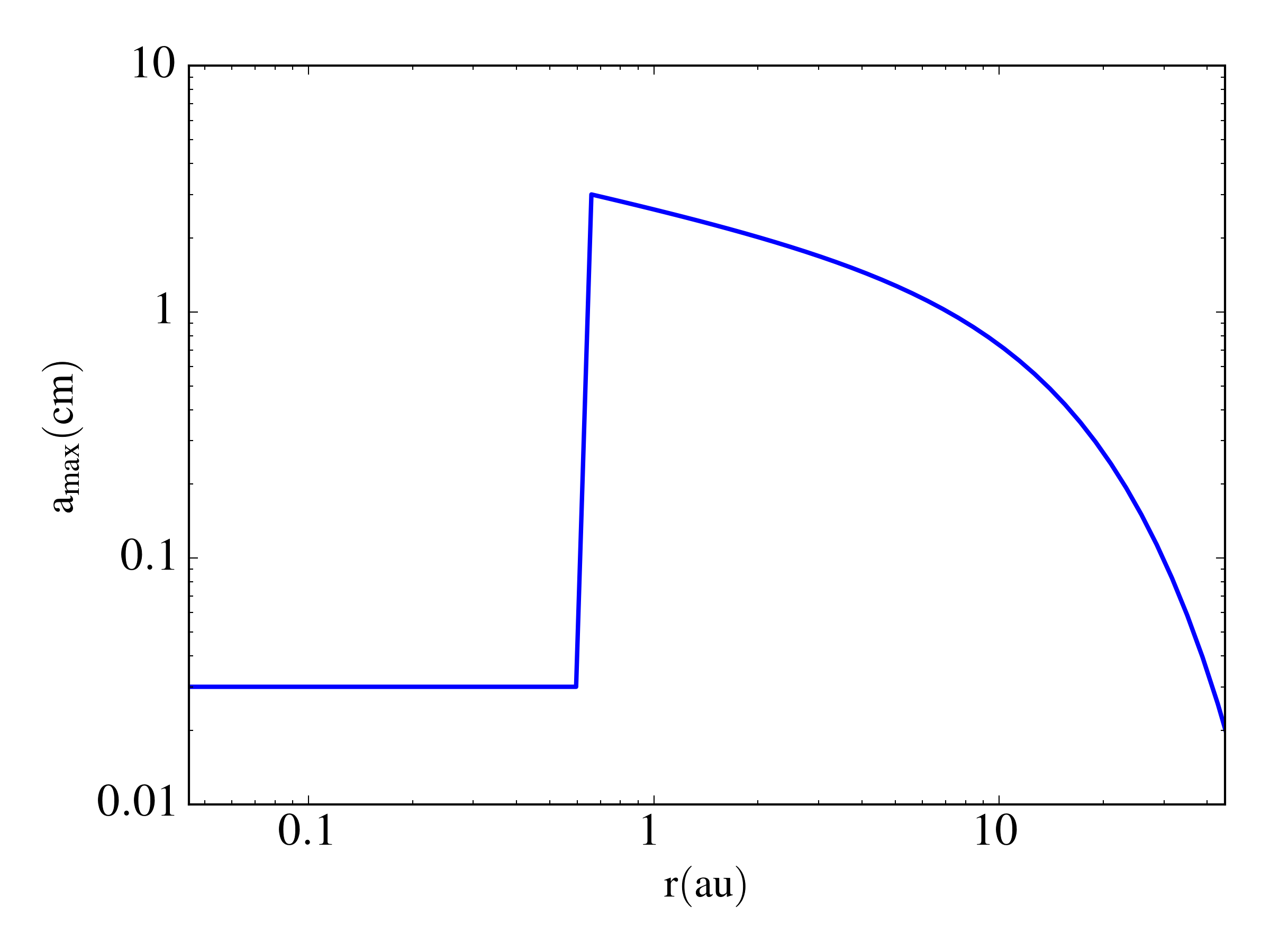}
			\caption{\textit{Top:} Schematic depiction of the behaviour of dust at the water snowline.  Dust grains beyond the water snowline possess large icy mantles which increase the fragmentation velocity, leading to enhanced grain growth and inward drift.  Dust grains interior to the snowline possess no ice mantles, are prone to fragmentation, and are thus are unable to grow as quickly to larger sizes. \textit{Left:} Surface density of solids as a function of radius assumed in our `migration' scenario, with a steep drop-off due to the H$_{2}$O snowline at 0.6\,au.  \textit{Right:} Maximum grain size as a function of radius assumed in our `migration' scenario, with a steep jump due to the H$_{2}$O snowline at 0.6\,au.}
			\label{fig:sigma_and_amax}
		\end{figure*}

		\subsection{The origin of the truncated disc: truncation or dust migration?}
		\label{subsec:origin_migration}
		
		Having shown that the observed SED is consistent with an optically thick dust disc with radius 0.6\,au, we enquire into possible origins for such a configuration.  
		One possibility is that the disc has been truncated by a dynamical encounter or due to the presence of a companion. 
		 A truncation radius of 0.6\,au would imply an encounter within a few au which would be extremely unlikely only considering encounters within the star forming environment, even in the densest star forming regions, let alone the rather sparse environment in which \xray\ is situated. A dynamical origin for this compact disc would instead need to invoke interactions occurring between bound companions in a multiple system. \protect\cite{StamatellosWhitworth2009} found that this mechanism could account for a wide range of disc radii: very small discs,  as in \xray\ would not however have been resolved in their simulations and would have been recorded as disc-less outcomes.

		\smallskip
		
		In the following subsection, we explore whether the steep decline in the SED can also be described in terms of dust radial drift. In particular we note that the temperature at the outer edge of the dust disc is suggestively close to that at the water snowline ($T\sim150\,$K) and we therefore explore whether the observed spectrum is consistent with a discontinuous change in grain properties associated with this snowline.
		
		\smallskip
		
		Outside of the snowline, water is frozen out into ice mantles on the grains.
		Therefore, grains do not fragment there as easily and can grow to large sizes \protect\citep{GundlachBlum2015}. 
		This in turns means that their dynamics progressively decouple from the influence of the gas and the grains thus drift inwards radially. Once they reach the water snowline, the ice mantles are sublimated; this leaves the grains prone to fragmentation, which decreases their sizes. This decrease in grain sizes causes the grains' dynamics to be more closely coupled to the gas which slows their radial drift.  Consequently the surface density of dust should {\it rise} interior to the water snowline. This process is depicted schematically in Figure~\ref{fig:sigma_and_amax}, top \citep[see also][]{BanzattiEtAl2015}. \protect\cite{OrmelEtAl2017}
		discuss this mechanism as a way of concentrating dust and making the TRAPPIST-1 system. We will henceforth refer to this model as the `migration' scenario.  
		
		\smallskip
		
		In Appendix~\ref{app:dust} we quantify the  jump in dust surface density and grain size that is to be expected if the fragmentation velocity changes discontinuously. We also derive expressions for the expected profile of maximum grain size and of dust surface density exterior to the water snowline on the assumption that grain growth in this region is limited by fragmentation and that the dust distribution is in a steady state (i.e. the radial mass flux is independent of radius). We show that while the dust radial surface density distribution scales as $r^{-1.5}$ (Figure \ref{fig:sigma_and_amax}, left), the maximum grain size distribution also depends on the radial gas profile. Figure \ref{fig:sigma_and_amax} (right) depicts the assumed variation of $a_{\rm max}$ with radius using Equations \ref{eqn:amax_vfrag} \& \ref{eqn:amax_radial}, in which the latter uses the radial gas profile of the form given in Equation \ref{eq:Sigmagas}.

		\smallskip

		The water snowline in this disc coincides well with the location at which we have so far assumed the radial truncation of the dust disc. In order to model whether the observed SED can also be explained with radial drift and fragmentation at the water snowline, we implement the effects on $a_\text{max}$ and $\Sigma_\text{dust}$ into DALI. In detail,  within $0.6$\,au we employ a single distribution of grain sizes with size distribution index $q=3$ in the range $5$nm to the local $a_\text{max}$  value but split the population at $10\,\mu$m so that small grains follow the gas while larger grains are vertically settled.  Outside of $0.6$\,au we use  a grain population (also with $q=3$, appropriate for an evolved grain population, see e.g. \protect\cite{BirnstielEtAl2012}) that extends in size from $10\,\mu$m to the local $a_\text{max}$ value. We treat this entire population as being settled (on the grounds that the large $a_\text{max}$ values and low surface densities in any case imply  negligible emission from small grains in this region).

		Although we have argued for the form of the $a_\text{max}$ and $\Sigma_\text{dust}$  distributions given in Equations \ref{eqn:amax_radial} and \ref{eqn:sigma_dust}, there are three further parameters that are required to specify the model, i.e. the over-all normalisations of the dust surface density distribution and $a_\text{max}$  distributions together with the outermost  radius of the dust disc. Evidently, if we make the outer radius sufficiently small then we can match the SED with a range of parameters since it is relatively easy to hide the spectral signatures of large dust particles  if they are only present over a limited radial zone. It is of more interest --- in order to contrast with the scenario presented in Section \ref{subsec:extent} where there is {\it no} dust beyond $0.6$\,au ---  to see if there are any viable models which fit the SED and where the dust is radially extended. Figure~\ref{fig:sigma_and_amax} exemplifies such a solution.  
		
		\smallskip
		
\begin{figure}
			\centering
			\includegraphics[width=\columnwidth]{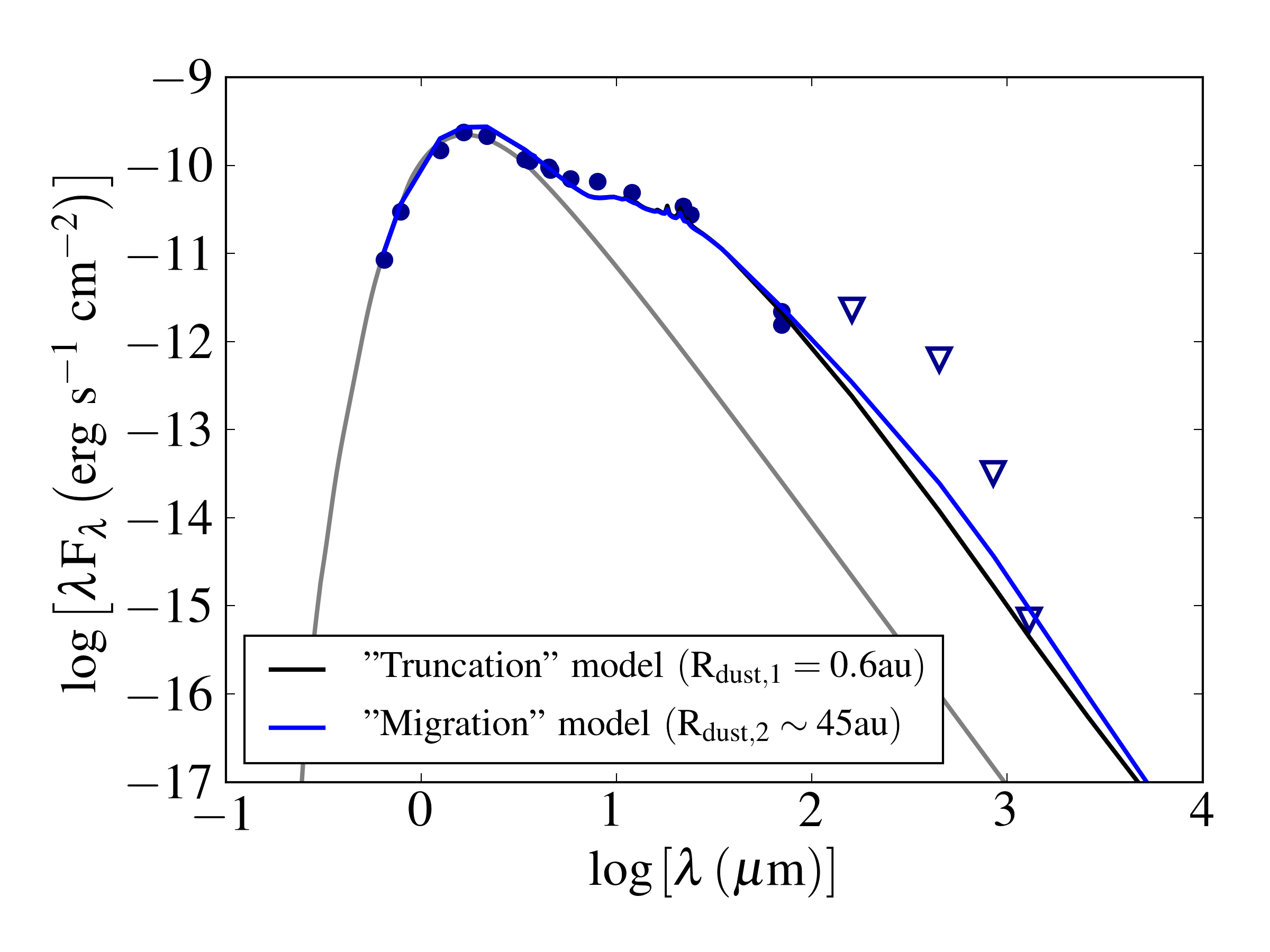}
			\caption{SED of the truncated model and the one with dust outside of 0.6\,au: The black thick line is the truncated model with $R_\text{dust, 1}=0.6\,$au and no dust outside of this radius. The blue model has $R_\text{dust, 1}=0.6\,$au, but $R_\text{dust, 2}\sim45\,$au. Both of them comply with the observed upper limit at 1.3\,mm.}
			\label{fig:SED_large}
		\end{figure}

		The values of  $a_\text{max}$ and dust surface density just outside the water snowline (3\,cm and $0.1$\,g cm$^{-2}$)  provide an acceptable fit to the SED even if the dust extends to $45$\,au (see Figure~\ref{fig:SED_large}). Smaller values of $a_\text{max}$ or larger dust surface density normalisation over-predict the flux at large wavelengths (especially the mm) since both these trends lead to increased optical depth at mm wavelengths. Models in this category would require a lower outer dust radius in order to be compatible with the SED. On the other hand, larger values of $a_\text{max}$ and lower values of dust surface density normalisation instead lower the optical depth in the near infrared to the point that they  under-predict the SED in this region and can therefore be ruled out. 
		
		\smallskip  
		
		It is worth emphasising that the `successful' model with an extended belt  of large grains (Figure~\ref{fig:SED_large}) is not necessarily a self-consistent outcome of dust growth models, since it is not guaranteed that the necessary value of $a_\text{max}$ and dust surface density will be simultaneously achieved and that dust growth will be fragmentation limited at this point (as we have assumed in our derivations in Appendix \ref{app:dust}).  Examining the properties of the  `successful' model, the dust surface density profile shown in Figure~\ref{fig:sigma_and_amax} (left) corresponds to total dust masses of $M_\text{dust}(\text{R}<0.6\text{au}) \sim 1 M_\oplus$ and $M_\text{dust}(\text{R}>0.6\text{au}) \sim 0.2 M_\oplus$.\footnote{Whereas the maximum dust mass inside of 0.6\,au is unbounded in the truncation scenario, the interior dust mass in the migration scenario needs to be fine tuned to about $1 M_\oplus$ in order for the large grains outside the H$_2$O snowline not to over-predict the mm flux. This dependence arises as the dust mass inside and outside of 0.6\,au are linked by the relationships plotted in Figure~\ref{fig:sigma_and_amax}. }
		It is only possible to constrain the required {\it gas} masses corresponding to the $a_\text{max}$ distribution shown in Figure \ref{fig:sigma_and_amax} (right) if one specifies the value of $\alpha$ \citep{Shakura1973} that controls the strength of turbulence and the value of the fragmentation velocity, $v_\text{frag}$.From Equation \ref{eq:Deltav} and \ref{eq:amax} the value of $\Sigma_\text{gas}$ corresponding to a particular grain size distribution scales as $\alpha/v_\text{frag}^2$; for $v_\text{frag} = 5$m s$^{-1}$ outside $0.6$\,au (and therefore and $v_\text{frag} = 0.5$m s$^{-1}$ inside $0.6$\,au) and $\alpha = 10^{-3}$ the gas mass interior and exterior to $0.6$\,au is around $40 M_\oplus$ and $1200M_\oplus$ respectively. However, given uncertainties in both $\alpha$ and $v_\text{frag}$ the constraints on the {\it gas} mass from this modelling are very weak.

		\subsection{Gas radius as discriminant between scenarios}
		
		So far we have considered two situations for the disc around \xray\ which would reproduce the observed SED (see Figure~\ref{fig:SED_large}).  The first involves a `truncation' scenario in which the dust is entirely confined within $0.6$\,au (for example due to interaction with a companion).  The second involves a `migration' scenario in which a combination of radial drift outside (and fragmentation inside) the H$_{2}$O snowline produces an optically thick inner dust disc and a low density region of large dust extending over many tens of au.  Whatever the cause, it is clear that SED modelling alone cannot distinguish between scenarios.
		
				\begin{figure*}
					\centering
					\includegraphics[width=0.46\textwidth]{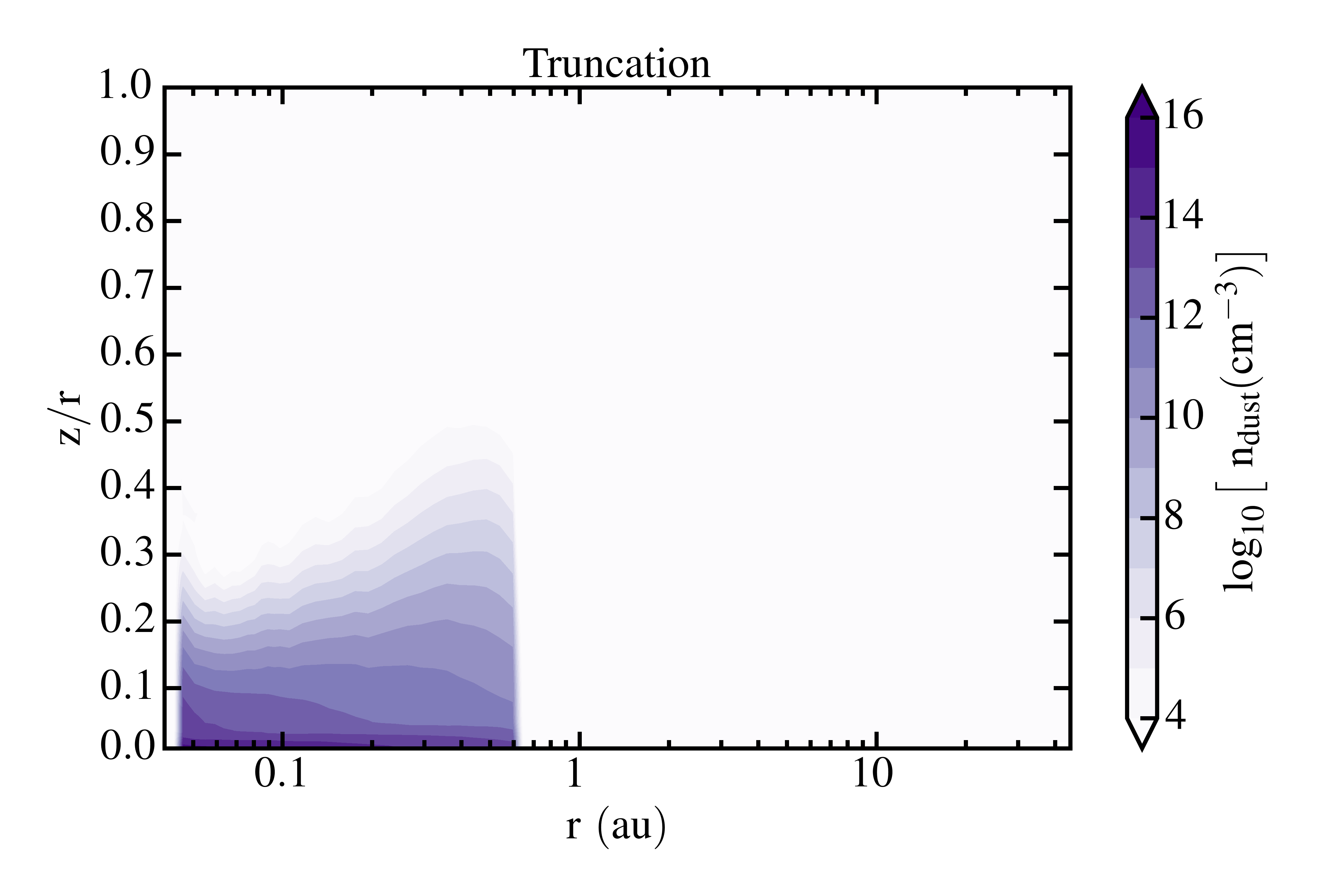}
					\hfill
					\includegraphics[width=0.46\textwidth]{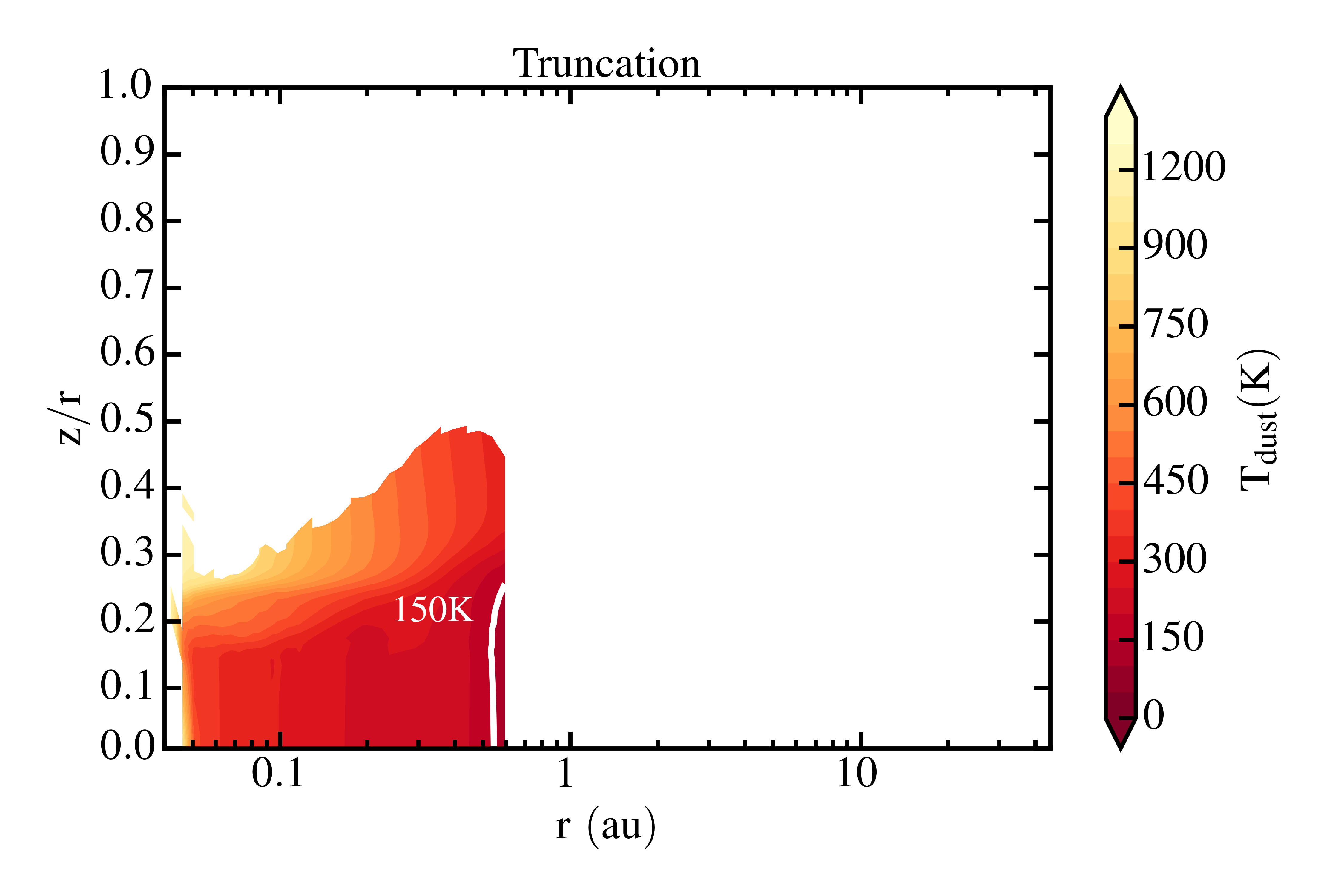}
					\includegraphics[width=0.46\textwidth]{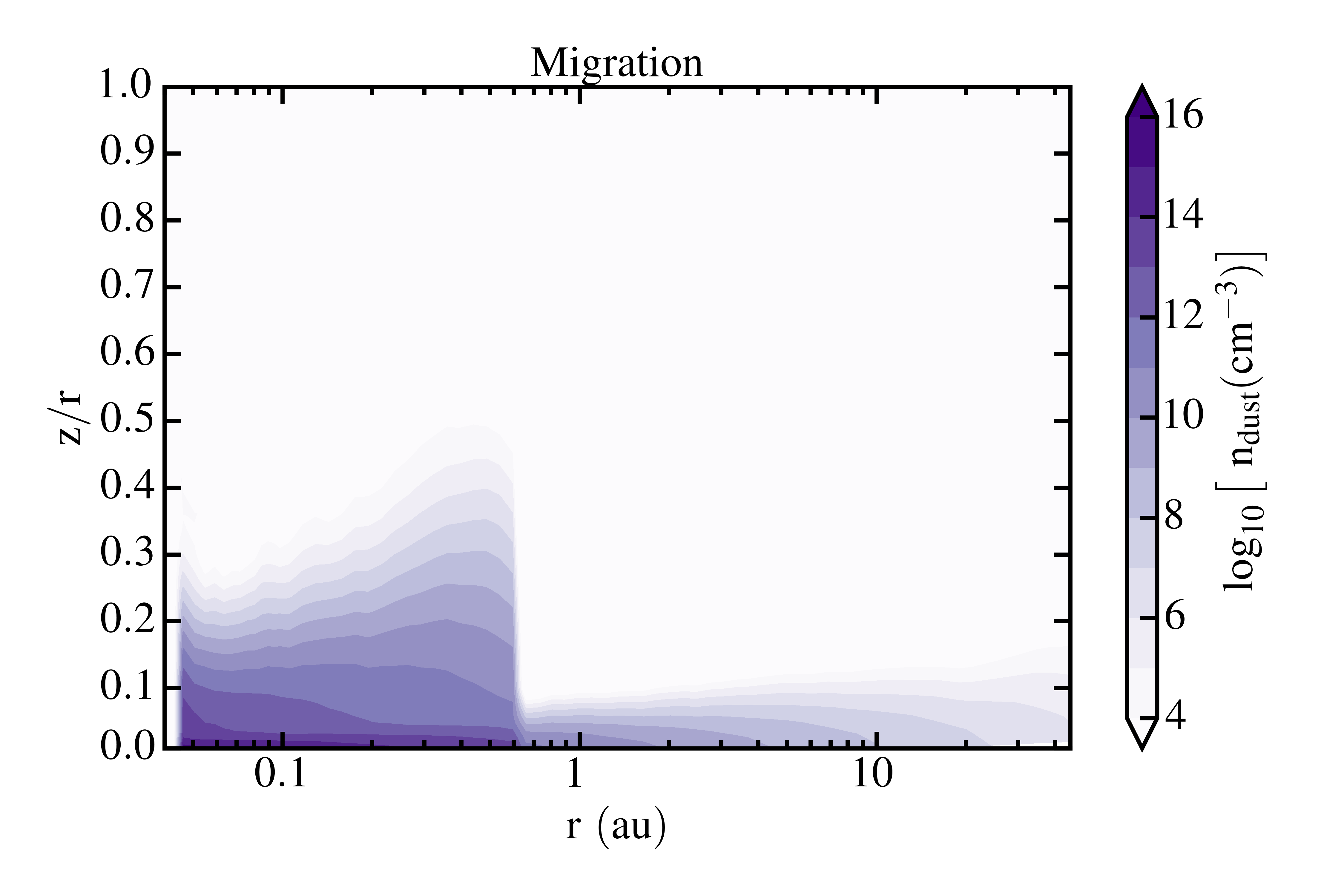}
					\hfill
					\includegraphics[width=0.46\textwidth]{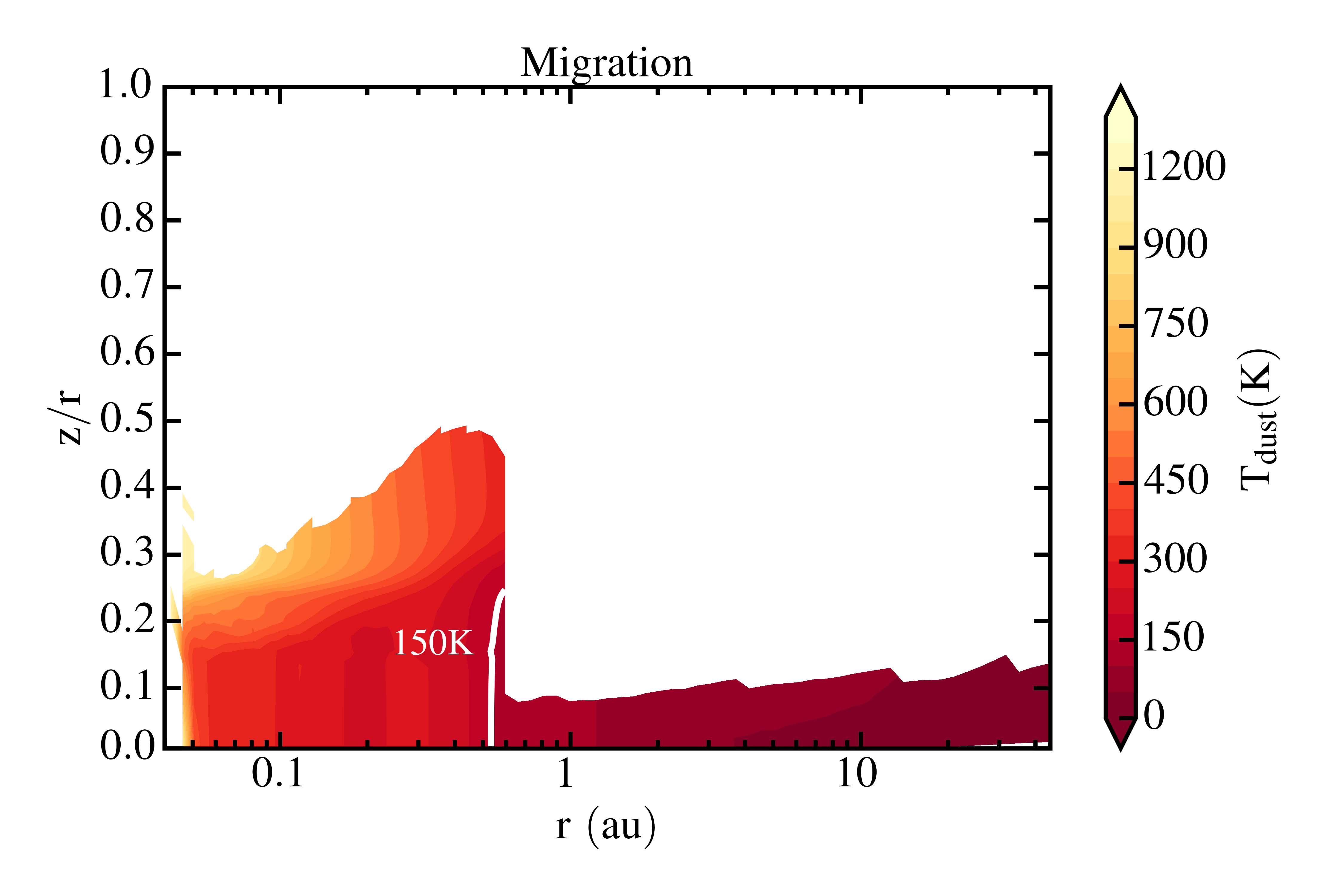}
					\caption{Dust density (left) and dust temperature (right) structure of the truncation (top) and migration (bottom) models.  In the dust temperature panels, contours of 150\,K are marked.}
					\label{fig:dens_temp_dust}
					
					\vfill
					
					\includegraphics[width=0.46\textwidth]{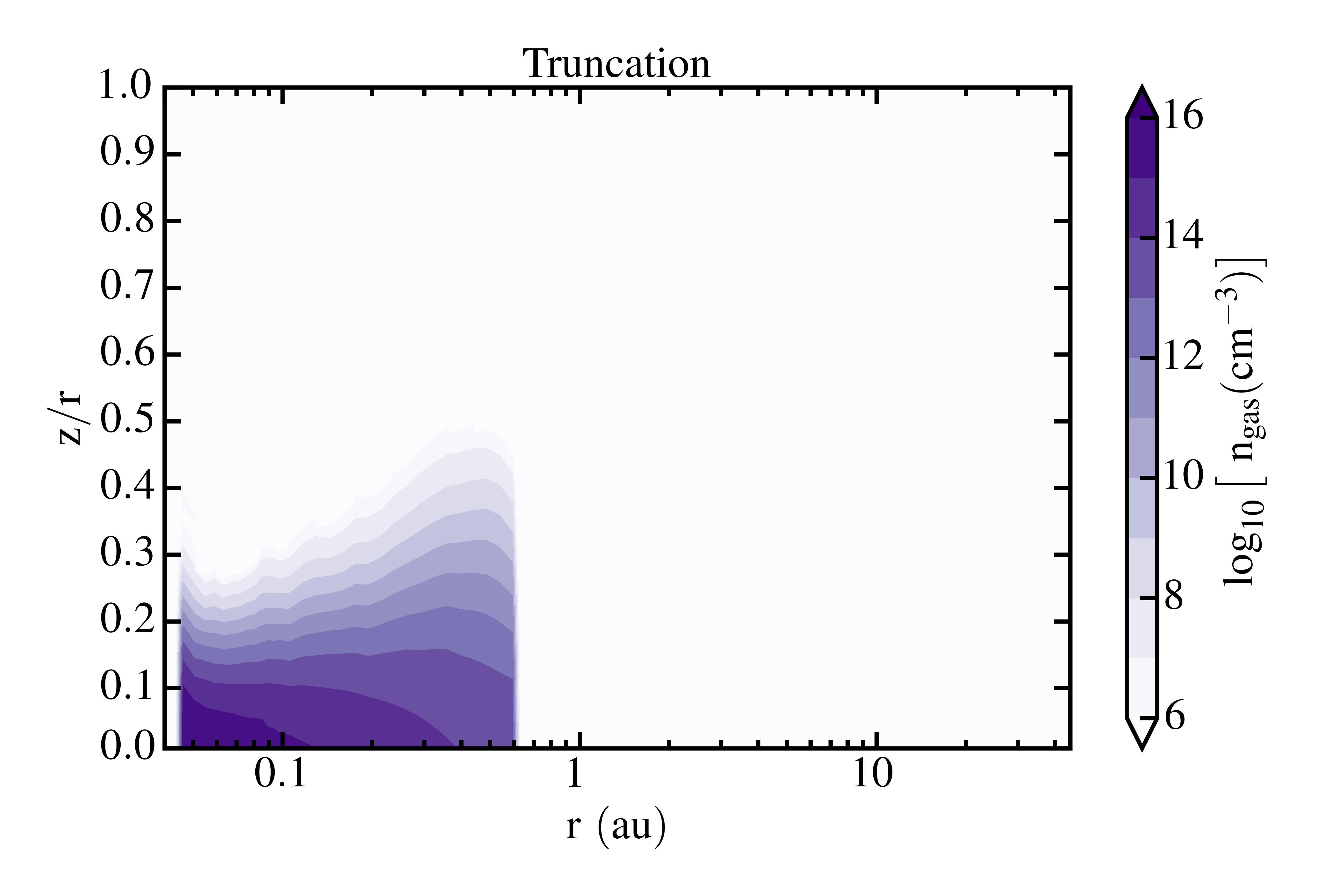}
					\hfill
					\includegraphics[width=0.46\textwidth]{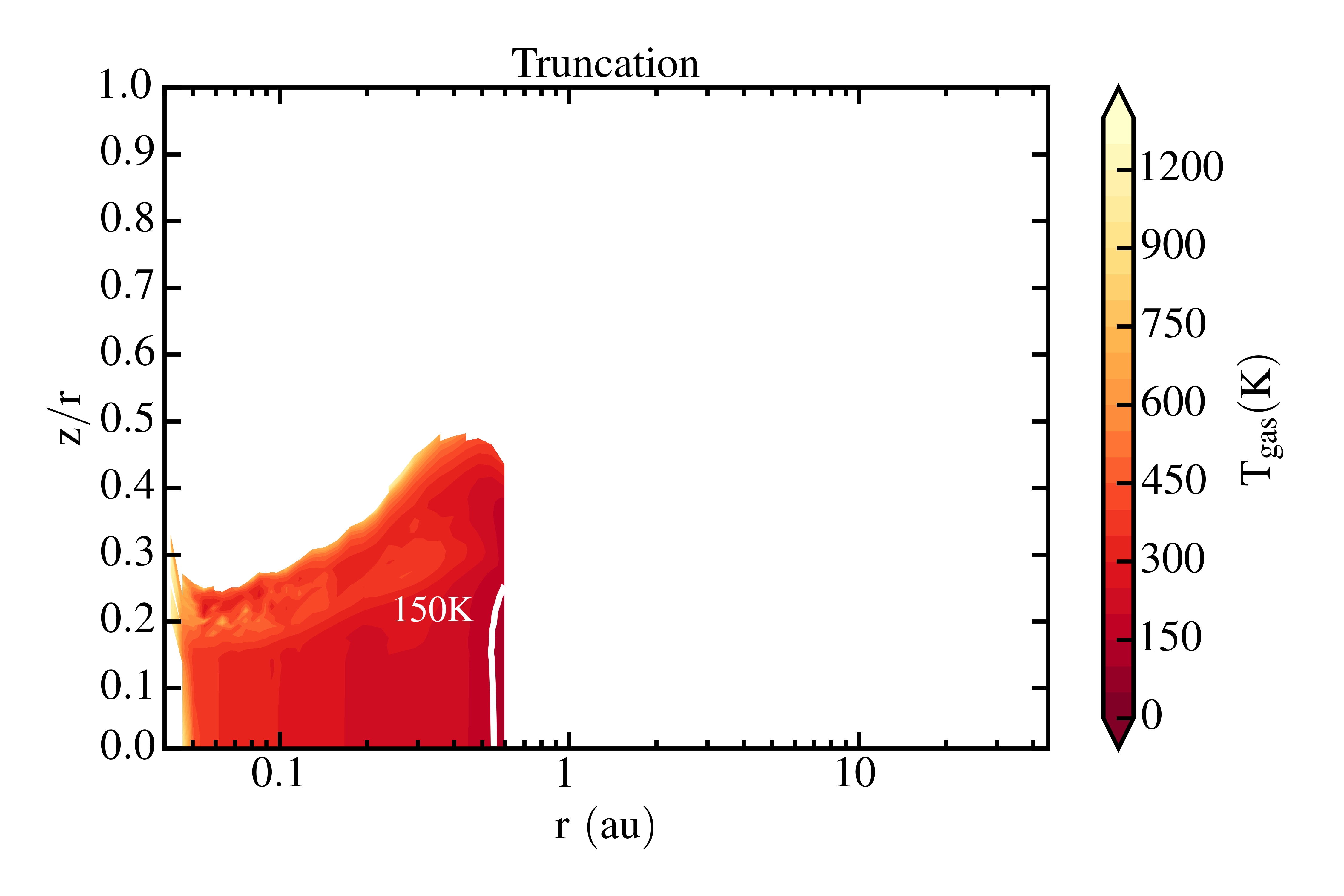}
					\includegraphics[width=0.46\textwidth]{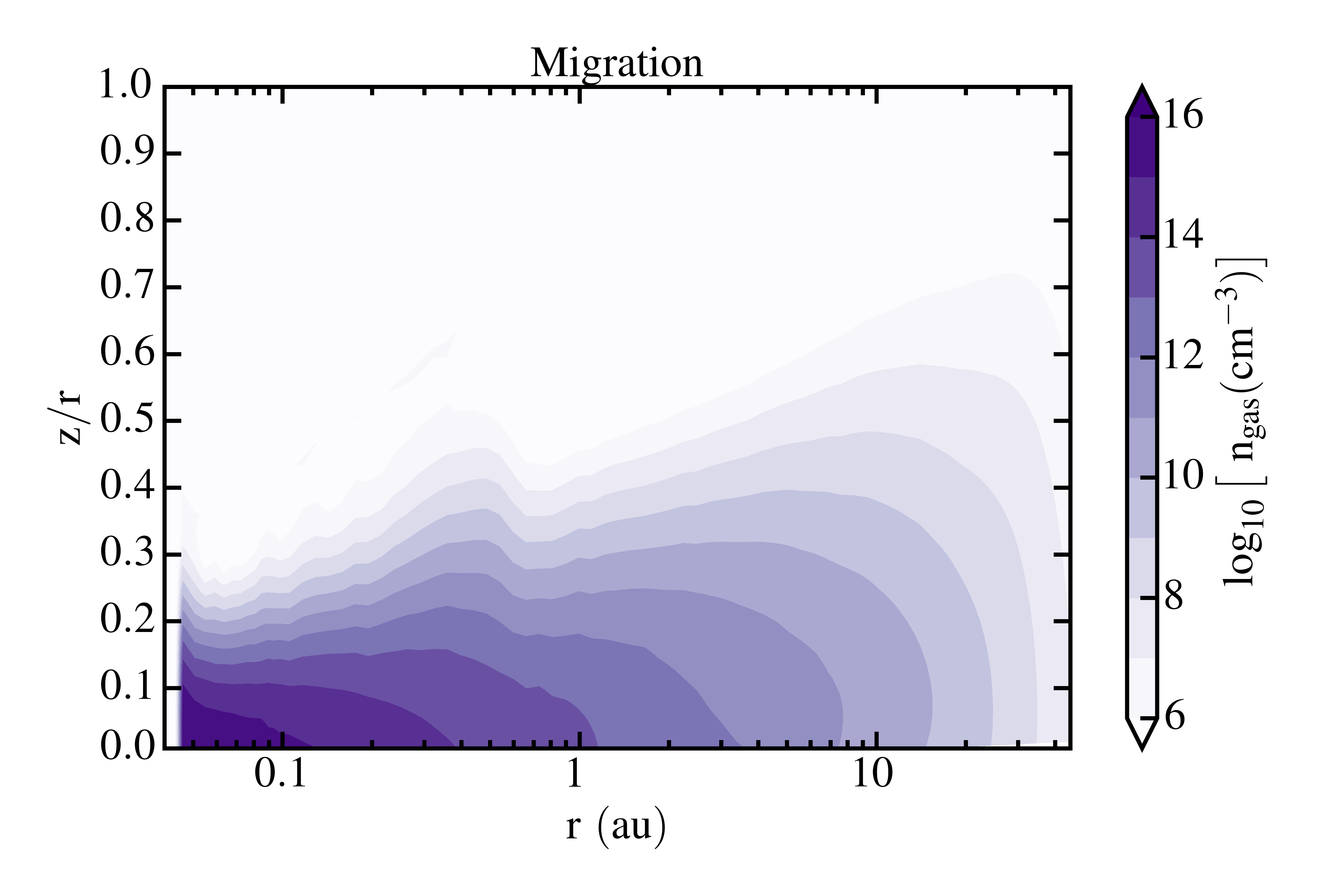}
					\hfill
					\includegraphics[width=0.46\textwidth]{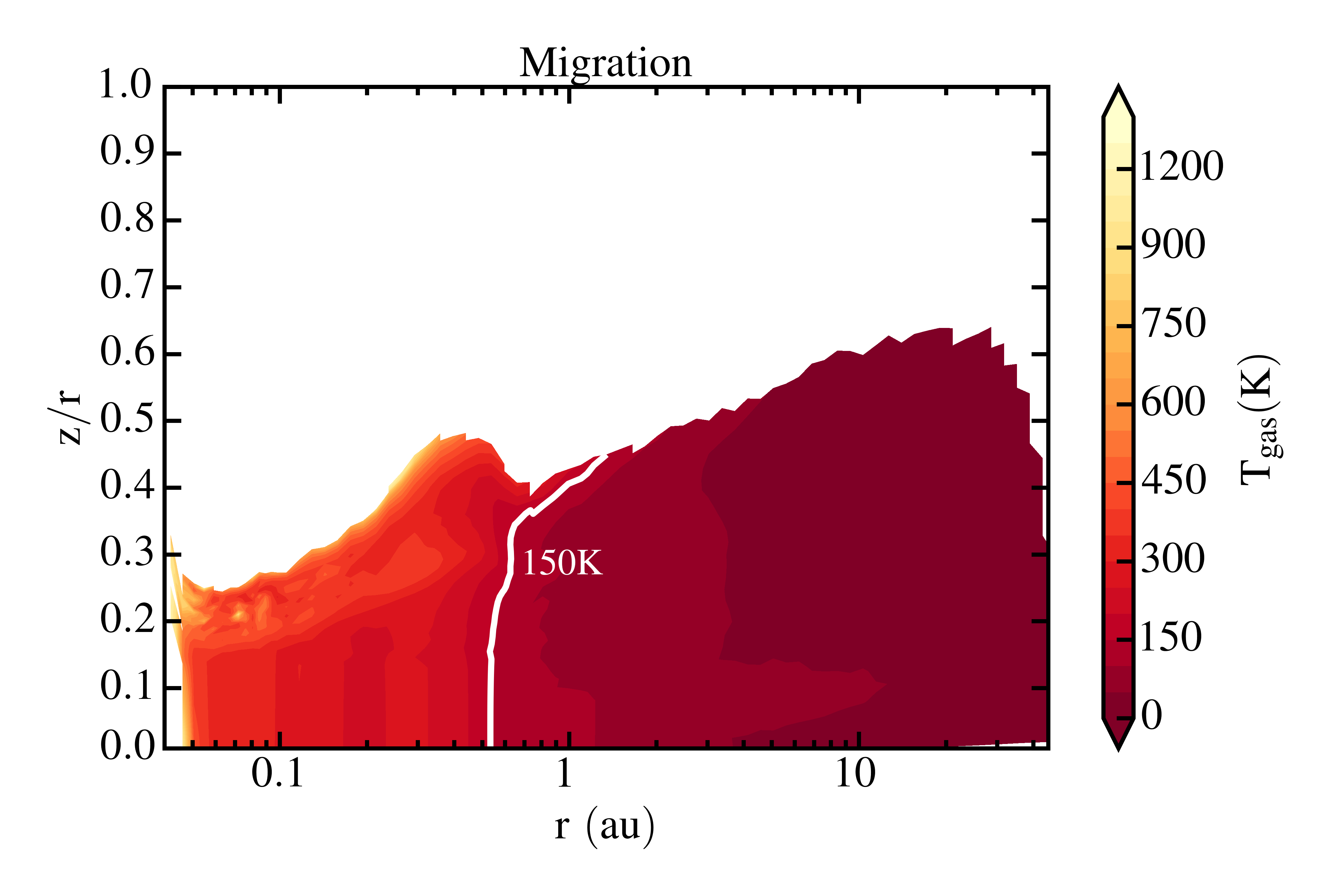}
					\caption{Gas density (left) and gas temperature (right) structure of the truncation (top) and migration (bottom) models.  In the gas temperature panels, contours of 150\,K are marked. The bump in the gas density plot of the `migration' scenario outside of 0.6\,au is an effect of plotting $z/r$ rather than $z$ on the y-axis.}
					\label{fig:dens_temp_gas}
				\end{figure*}
		
		\smallskip
		
		A clear discriminating difference between these two scenarios would be the radial extent of the {\it gas} in the disc -- in the former the gas would have a similarly small radius as the dust, while in the latter the gas disc would share a similarly extended configuration to the large dust grains.\footnote{Although the gas radius may even exceed the radius of the large grains \protect\citep[see][]{BirnstielAndrews2014}.}  As such, observations of gas tracers towards \xray\ may have the ability to determine which configuration is at work.  Figure \ref{fig:dens_temp_gas} shows the corresponding gas density and temperature structure for the two scenarios we consider.

		\smallskip
		
		In both scenarios, the dust interior to 0.6\,au intercepts the stellar irradiation and causes flared structure to appear in the gas, typical of vertical hydrostatic equilibrium models \citep[see, e.g.,][]{Woitke2009}.  The large dust in the `migration' scenario is extremely settled, as such has little effect on the gas temperature in these regions. In these midplane regions, gas and dust temperatures are tightly coupled and in thermal equilibrium due to the high densities \citep[e.g.][]{2017A&A...605A..16F}, as can be seen by the relatively invariant temperature contours.   
		
		\smallskip

		\begin{figure}
			\centering
			\includegraphics[width=0.48\textwidth]{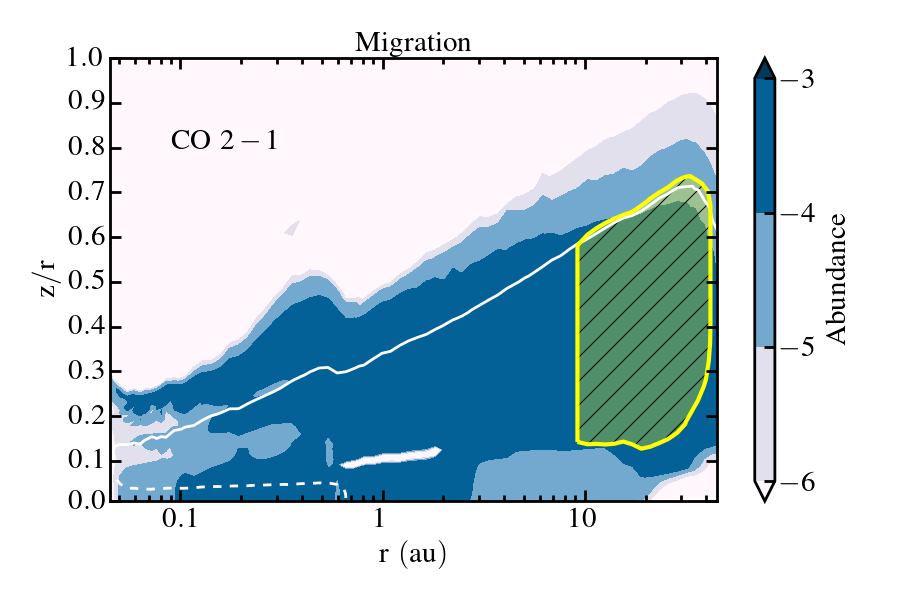}
			\caption{Fractional abundance of CO throughout a disc model in the `migration' case with $R_\text{gas}\sim30\,$au (colour scale), overlaid with the region which is responsible for 75 per cent of the emission for the $J = 2$ -- 1 transition (yellow hatch).  The white solid line denotes the $\tau=1$ surface for the $J = 2$ -- 1 transition, while the white dashed line indicates the $\tau=1$ surface for the dust at the wavelengths of this transition (i.e. 1.3\,mm).}
			\label{fig:cbf_drift_trunc}
		\end{figure}
		
		The combined thermochemical nature of the DALI code allows us to easily examine the line emission from these scenarios.  In order to determine the extent of the gas in the disc, we need to consider lines that may be emitted from large radii. Given the faintness of the source and lack of any surrounding cloud material, we have chosen to examine the observability of the $^{12}$CO isotopologue (hereafter referred to as CO).  Figure \ref{fig:cbf_drift_trunc} shows the abundance of CO throughout the disc for the `migration' case, where the gas disc extends to $\sim30$\,au. This model was calculated using a time-dependent chemical solver, but models calculated with time-independent chemistry yield the same CO $J = 2$ -- 1 line fluxes.
		The highest abundances are seen in a warm molecular layer in the inner disc at relative heights of $z/r \sim 0.2$ -- 0.4, and throughout the entirety of the disc exterior to 0.6\,au.  Disc layers above this are subject to significant amounts of UV irradiation which act to photodissociate the CO molecule. In the inner disc midplane regions with high temperatures, C is preferentially in CH$_{4}$, whereas in the colder midplane regions further out, CO will be depleted due to freeze-out (where dust grains are present). 
		Despite the relative ubiquity of CO throughout this disc, the majority (75 per cent) of the emission of the $J = 2$ -- 1 transition is confined to radii between 10 -- 30\,au at a relative height of between 0.15 -- 0.7 (yellow hatched region).  From inspection of the optical depth surfaces shown in Figure \ref{fig:cbf_drift_trunc}, it is clear that the CO emission originates far above the $\tau=1$ surface for the dust at 1.3\,mm (dotted line), but is beneath the $\tau=1$ surface for the line (solid line), and is thus optically thick.  Therefore, the CO $J=2$--1 emission is a useful proxy of the outer gas disc radius. Nevertheless, Figure \ref{fig:cbf_drift_trunc} only depicts the results of a single model.
		The presence of large dust grains or the lack thereof outside of 0.6\,au has essentially no effect on the CO abundance or temperature in the region where 75 per cent of the emission comes from. The integrated CO $J = 2$ -- 1 line fluxes of both cases are therefore essentially identical. 
		The reason is that the additional dust outside of 0.6\,au is very settled and optically thin and therefore has a negligible effect on the temperature structure of the disc.  
		
		\smallskip
		
\begin{figure}
			\centering
			\includegraphics[width=\columnwidth]{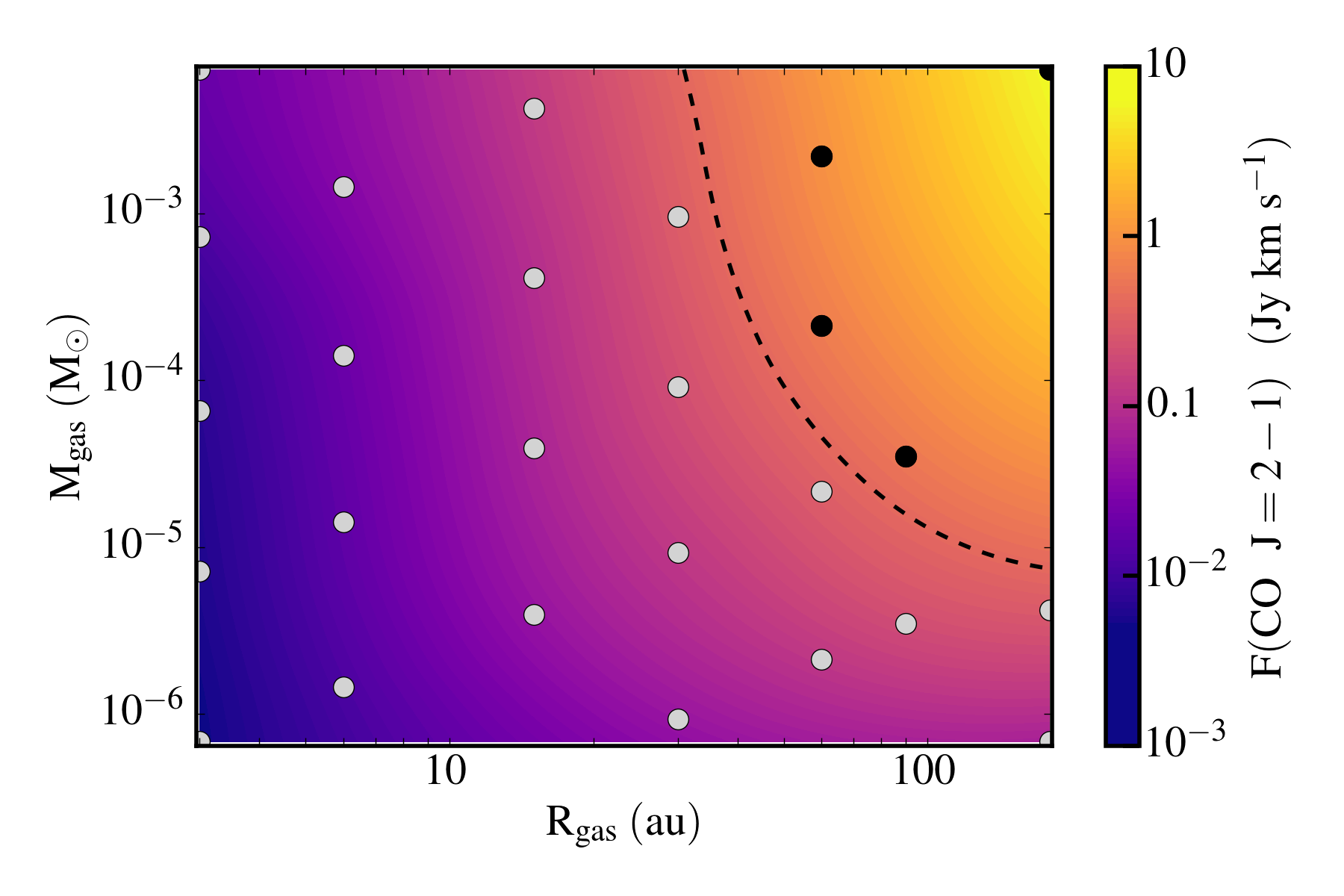}
			\caption{Resulting $^{12}$CO $J=2$--1 fluxes of our grid of models (points) in which we vary the gas radius and the gas mass.  Values between these points are calculated using cubic interpolation on the triangular grid connecting each point.  The dashed line indicates a flux of 0.4 Jy km\,s$^{-1}$, the upper limit for the $^{12}$CO $J=2$--1 derived in Section~\ref{sec:observations}.  Black points indicate models which violate this upper limit.}
			\label{fig:flux}
		\end{figure}

		To investigate a larger parameter space, we have run a suite of 26 models.  Each model produces a fit to the observed SED, but the gas radius  $R_\text{c}$ and the total gas mass $M_\text{gas}$ are varied.  All other stellar and disc parameters are kept fixed to the values listed in Table~\ref{tab:prop}.  Our chosen ranges of gas mass are set by the requirement that the disc must obey gravitational stability (and thus M$_{\rm gas} < 0.1$M$_{\rm star}$), but must contain sufficient dust mass to reproduce the observed $70 \mu$m flux in the SED.  Our chosen ranges of gas radius span approximately 3 to 200\,au.  Figure \ref{fig:flux} shows the resulting CO $J = 2$ -- 1 line fluxes obtained from the models within the grid. Again,the presence of the very large grains outside of 0.6\,au (or their lack) has no effect on the total CO fluxes in our models.
		
		\smallskip
		
		As can be seen, there is a weak dependence between the resulting $J=2$--1 line flux and the value of $M_{\rm gas}$ assumed for gas masses greater than $\approx 10^{-5}$\,M$_{\odot}$, due to the fact that the emission is largely optically thick in this regime.  For gas masses lower than $\approx 10^{-5}$\,M$_{\odot}$, a somewhat stronger dependence starts to emerge.  However, by far the strongest dependence is between the $J=2$--1 line flux and the value of $R_{\rm gas}$.  Such a result is not surprising, because due to the optically thick nature of the line emission, the line flux will simply scale geometrically with the emitting area.  Though we do not possess a firm detection of the $J=2$--1 line flux toward \xray, a combination of our calculated non-detection of $<0.4$\,Jy\,km\,s$^{-1}$ and our model grid already allows us to exclude several configurations.  These broadly consist of discs with an outer gas radius greater than 50\,au and a gas mass greater than $10^{-4}$\,M$_{\odot}$.  
		Other studies of classical T Tauri Stars have used the $^{12}$CO total flux (or upper limit) to constrain the radial extent of gaseous disks \protect\citep[e.g.][]{WoitkeEtAl2011} where spatially resolved observations were not available.
		Given the weak dependence on M$_{\rm gas}$, but stronger dependence on R$_{\rm gas}$, a firm detection (or more stringent non-detection) would likely lead the gas radius to be determined rather unambiguously, possibly allowing the 'truncation' or 'migration' scenarios to be distinguished.

		\section{Conclusions}
		\label{sec:conclusion}
		
		We have modelled the disc around the very low mass star \xray. Including an upper limit on the 1.3\,mm continuum flux from recent ALMA observations gives an additional constraint on the disc structure. We find that its SED can be explained by a very truncated, optically thick dust component with $R_\text{dust, 1}=0.6\,$au.
		We have explored two scenarios that can potentially explain this finding. On the one hand, dynamical `truncation' of the disc could have shrunk the dust disc to its tiny radius. On the other hand, the SED is also compatible with some amount of large dust outside of 0.6\,au. For this scenario, radial `migration' coupled with fragmentation interior to the H$_{2}$O snowline (located at 0.6\,au) result in  a cliff in $\Sigma_\text{dust}$ of two orders of magnitude and in a sudden increase of grain sizes. Our modelling shows that this scenario matches the observed SED for outer dust disc radii of $R_\text{dust, 2}\lesssim 45\,$au.
		
		\smallskip
		
		In both scenarios, a dust mass inside of 0.6\,au bigger than $\sim1 M_\oplus$ is required to match the observed SED, however due to the optically thick nature of the dust, we do not obtain an upper limit on the dust mass. In the `migration' scenario, the dust mass needs to be fine-tuned to $\sim1 M_\oplus$ inside the H$_2$O water snowline in order to match the SED as the dust masses inside and outside of 0.6\,au are dependent on each other.
		
		\smallskip
		
		In order to distinguish further between these scenarios, the gas radius of the disc can offer important information. If it is truncated at very small radii as well, a dynamical truncation of the disc seems likely. If it is more extended, radial drift in combination with dust physics at the water snowline offer a plausible explanation. We present predictions for CO $J=2$--1 fluxes, invoking various $M_\text{gas} - R_\text{gas}$ combinations, that will enable future observations to potentially distinguish between these two scenarios.  
		
		\smallskip
		
		Finally, we note that regardless of the underlying mechanism for the small scale dust disc, its optically thick nature suggests that several $M_\oplus$ of dust can be located within a small radial extent from the central star.  Such a configuration suggests that \xray\ may be a potential precursor to a system of tightly packed terrestrial planets, such as the recently discovered TRAPPIST-1 system.


		\section*{Acknowledgements}
		We would like to thank the anonymous reviewer for their thorough report and suggestions that have improved the work.
		We thank Luca Matr{\`a}, Attila Juh{\'a}sz, Mihkel Kama, Leonardo Testi and Ewine van Dishoeck for helpful discussions.  This work has been supported by the DISCSIM project, grant agreement 341137 funded by the European Research Council under ERC-2013-ADG. DMB is funded by this ERC grant and an STFC studentship.  This  paper  makes  use  of  the  following  ALMA  data: ADS/JAO.ALMA\#2013.1.00163.S. ALMA is a partnership of ESO (representing its member states), NSF (USA) and NINS (Japan), together with  NRC (Canada),  NSC and  ASIAA  (Taiwan), and  KASI (Republic of Korea), in cooperation with the Republic of Chile. The Joint ALMA Observatory is operated by ESO,  AUI/NRAO and NAOJ.

		
		
		
		\bibliographystyle{mnras}
		\bibliography{bibliography.bib} 

		
		
		
		\appendix
		
		\section{Dust processing at the water snowline}
		\label{app:dust}
		In Section~\ref{subsec:origin_migration}, we study whether the truncated dust disc could be related to physical processes happening at the water snowline.  The key property that changes across the snowline is $v_\text{frag}$, the maximum collision velocity of grains for which coagulation, rather than fragmentation, is to be expected. Following \cite{BlumWurm2000,GundlachBlum2015}, $v_\text{frag}$ is reduced by an order of magnitude interior to the snowline. If grain growth is limited by fragmentation, the local value of $v_\text{frag}$ is linked to the maximum expected grain size through equating the relative collision velocity $\Delta v$ for the largest grains at any location with $v_\text{frag}$. $\Delta v$ for grains that are coupled to a turbulent flow via  drag forces is given by \protect\citep{VoelkEtAl1980,OrmelCuzzi2007}:
		\begin{equation}
			\Delta v\approx\sqrt{\alpha~\rm{St}}~c_\text{s} \hspace{2pt},
			\label{eq:Deltav}
		\end{equation}
		where $\alpha$ is the Shakura-Sunyaev viscosity parameter \citep{Shakura1973}, $c_\text{s}$ the sound speed and St is the local Stokes number (ratio of drag time to dynamical time) for the largest grains. 
		In the Epstein drag regime, the midplane St for grains of size $a_\text{max}$ is given by
		\begin{equation}
			a_\text{max}=\frac{2~\Sigma_\text{gas}~\rm{St}}{\pi \rho_\text{grain}} \hspace{2pt},
			\label{eq:amax}
		\end{equation}
		where we have assumed a Gaussian vertical structure.
		When growth is limited by fragmentation, the size of the largest grains (radius $a_\text{max}$) is set by $\Delta v \approx v_\text{frag}$.
		Combining Equations \ref{eq:Deltav} and \ref{eq:amax} then yields
		\begin{equation}
			\label{eqn:amax_vfrag}
			a_\text{max} \propto v_\text{frag}^2 \hspace{2pt}.
		\end{equation}
		This change in maximum grain size also affects the radial drift of the largest grains (which, for typical expected grain size distributions, dominate the local dust mass budget). Since the rate of radial drift for particles with St<1 scales as St we also expect
		\begin{equation}
			v_\text{drift} \propto v_\text{frag}^2 \hspace{2pt}.
		\end{equation}
		Continuity then implies that the local dust surface densities at an interface where $v_\text{frag}$ changes steeply scales as
		\begin{equation}
			\Sigma \propto v_\text{frag}^{-2}
		\end{equation}
		on each side of the interface. In summary, then, such a model predicts (if $v_\text{frag}$ varies by a factor 10 at the snowline) that the maximum grain size should be {\it reduced} by  two orders of magnitude within the snowline, while the total surface density of solids should {\it increase} by the same factor.
		
		In order to derive the radial dependence of $a_\text{max}$ outside of the water snowline in the fragmentation dominated regime,  we rearrange Equation~\ref{eq:Deltav} and invoke  the thin disc hydrostatic equilibrium relation $c_s = (H/r) v_\Phi$  so as to obtain the Stokes number of the largest grains: 
		\begin{equation}
			\label{eq:Stfrag}
			\text{St}_\text{frag}=\frac{0.37}{3\alpha}\left( \frac{v_\text{frag}}{v_\Phi} \right)^2 \left( \frac{r}{H} \right)^2 \hspace{2pt},
		\end{equation}
		where $v_\Phi=\sqrt{\frac{GM_\text{star}}{r}}$, and the numerical factors are  determined from fits to simulations \protect\citep{BirnstielEtAl2012}. Using $c_\text{s}^2 \propto r^{-0.5}$ and $\Sigma_\text{gas}$ from Equation~\ref{eq:Sigmagas} 
		we then obtain an expression for the radial dependence of the maximum grain size of the form: 
		\begin{equation}
			\label{eqn:amax_radial}
			a_\text{max}(r) \propto r^{-0.3} \exp{\left[- \left( \frac{r}{R_\text{c}}\right)^{1.2} \right]} \hspace{2pt}.
		\end{equation}
		In order to calculate the radial profile of the dust surface density we invoke a state of steady flow so that 
		\begin{equation}
			\Sigma_\text{dust} v_\text{drift} r=\rm{const} \hspace{2pt}.
		\end{equation}
		In the limit $St < 1$ the radial drift velocity is given by 
		\begin{equation}
			v_\text{drift}\sim \text{St}_\text{frag} \left( \frac{H}{r} \right)^2 v_\Phi 
		\end{equation}
		\protect\citep{TakeuchiLin2002} and substituting for $\text{St}_\text{frag}$ from Equation~\ref{eq:Stfrag} then yields 
		the radial dependence of $\Sigma_\text{dust}$  as 
		\begin{equation}
			\label{eqn:sigma_dust}
			\Sigma_\text{dust} \propto r^{-1.5} \hspace{2pt},
		\end{equation}
		where the exponential cut-off of the gas distribution cancels out.

		\bsp	
		\label{lastpage}
	\end{document}